\pdfoutput=1
\documentclass[12pt,a4paper]{article}

\usepackage{ifthen} 
\newboolean{pdflatex}
\setboolean{pdflatex}{true} 

\newboolean{articletitles}
\setboolean{articletitles}{true} 

\newboolean{uprightparticles}
\setboolean{uprightparticles}{false} 

\def\paperauthors{LHCb collaboration} 
\def\paperasciititle{Study of the Bose-Einstein correlations of identical pions in proton-lead collisions} 
\def\papertitle{Study of the Bose-Einstein correlations of same-sign pions in proton-lead collisions} 
\def\paperkeywords{{High Energy Physics}, {LHCb}} 
\def\papercopyright{\the\year\ CERN for the benefit of the LHCb collaboration} 
\def\paperlicence{CC BY 4.0 licence}
\def\paperlicenceurl{https://creativecommons.org/licenses/by/4.0/}

\usepackage[top=1in, bottom=1.25in, left=1in, right=1in]{geometry}

\usepackage{microtype}
\usepackage{lineno}  
\usepackage{xspace} 
\usepackage{caption} 

\usepackage{graphicx}  
\usepackage{color}
\usepackage{colortbl}
\graphicspath{{./figs/}} 

\usepackage{amsmath} 
\usepackage{amssymb}
\usepackage{amsfonts}
\usepackage{upgreek} 

\newcommand*\patchAmsMathEnvironmentForLineno[1]{%
\expandafter\let\csname old#1\expandafter\endcsname\csname #1\endcsname
\expandafter\let\csname oldend#1\expandafter\endcsname\csname
end#1\endcsname
 \renewenvironment{#1}%
   {\linenomath\csname old#1\endcsname}%
   {\csname oldend#1\endcsname\endlinenomath}%
}
\newcommand*\patchBothAmsMathEnvironmentsForLineno[1]{%
  \patchAmsMathEnvironmentForLineno{#1}%
  \patchAmsMathEnvironmentForLineno{#1*}%
}
\AtBeginDocument{%
\patchBothAmsMathEnvironmentsForLineno{equation}%
\patchBothAmsMathEnvironmentsForLineno{align}%
\patchBothAmsMathEnvironmentsForLineno{flalign}%
\patchBothAmsMathEnvironmentsForLineno{alignat}%
\patchBothAmsMathEnvironmentsForLineno{gather}%
\patchBothAmsMathEnvironmentsForLineno{multline}%
\patchBothAmsMathEnvironmentsForLineno{eqnarray}%
}

\usepackage{hyperxmp}

\usepackage[pdftex,
            pdfauthor={\paperauthors},
            pdftitle={\paperasciititle},
            pdfkeywords={\paperkeywords},
            pdfcopyright={Copyright (C) \papercopyright},
            pdflicenseurl={\paperlicenceurl}]{hyperref}

\usepackage[colorinlistoftodos,textsize=scriptsize]{todonotes}

\usepackage[bottom,flushmargin,hang,multiple]{footmisc}

\usepackage[all]{hypcap} 

\usepackage{xspace} 
\usepackage{upgreek}

\def\lhcb   {\mbox{LHCb}\xspace}

\def\cms    {\mbox{CMS}\xspace}

\def\lhc    {\mbox{LHC}\xspace}

\def\velo   {VELO\xspace}

\def\MagUp {\mbox{\em Mag\kern -0.05em Up}\xspace}

\ifthenelse{\boolean{uprightparticles}}%
{

 \def\Ppi         {\ensuremath{\uppi}\xspace}                 
                  
 \def\Prho        {\ensuremath{\uprho}\xspace}

 \def\PDelta      {\ensuremath{\Delta}\xspace}                 
 \def\PXi         {\ensuremath{\Xi}\xspace}                 
 \def\PLambda     {\ensuremath{\Lambda}\xspace}                 
 \def\PSigma      {\ensuremath{\Sigma}\xspace}                 
 \def\POmega      {\ensuremath{\Omega}\xspace}                 
 \def\PUpsilon    {\ensuremath{\Upsilon}\xspace}
 \let\oldPi\Pi
 \def\PPi         {\ensuremath{\oldPi}\xspace}

 \def\PB      {\ensuremath{\mathrm{B}}\xspace}                 
                  
 \def\PD      {\ensuremath{\mathrm{D}}\xspace}

 \def\PK      {\ensuremath{\mathrm{K}}\xspace}

 \def\Pe      {\ensuremath{\mathrm{e}}\xspace}

 \def\Pi      {\ensuremath{\mathrm{i}}\xspace}

 \def\Pp      {\ensuremath{\mathrm{p}}\xspace}

 \def\Ps      {\ensuremath{\mathrm{s}}\xspace}

 \def\thebaroffset{0.0em}
}
{

 \def\Ppi         {\ensuremath{\pi}\xspace}                 
                  
 \def\Prho        {\ensuremath{\rho}\xspace}

 \mathchardef\PDelta="7101
 \mathchardef\PXi="7104
 \mathchardef\PLambda="7103
 \mathchardef\PSigma="7106
 \mathchardef\POmega="710A
 \mathchardef\PUpsilon="7107
 \mathchardef\PPi="7105
                  
 \def\PB      {\ensuremath{B}\xspace}                 
                  
 \def\PD      {\ensuremath{D}\xspace}

 \def\PK      {\ensuremath{K}\xspace}

 \def\Pe      {\ensuremath{e}\xspace}

 \def\Pi      {\ensuremath{i}\xspace}

 \def\Pp      {\ensuremath{p}\xspace}

 \def\Ps      {\ensuremath{s}\xspace}

 \def\thebaroffset{0.18em}
}
\newcommand{\offsetoverline}[2][\thebaroffset]{\kern #1\overline{\kern -#1 #2}}%

\makeatletter
\ifcase \@ptsize \relax
  \newcommand{\miniscule}{\@setfontsize\miniscule{4}{5}}
\or
  \newcommand{\miniscule}{\@setfontsize\miniscule{5}{6}}
\or
  \newcommand{\miniscule}{\@setfontsize\miniscule{5}{6}}
\fi
\makeatother

\DeclareRobustCommand{\optbar}[1]{\shortstack{{\miniscule (\rule[.5ex]{1.25em}{.18mm})}
  \\ [-.7ex] $#1$}}

\def\epem       {{\ensuremath{\Pe^+\Pe^-}}\xspace}

\def\squark    {{\ensuremath{\Ps}}\xspace}

\def\pion   {{\ensuremath{\Ppi}}\xspace}

\def\pip    {{\ensuremath{\pion^+}}\xspace}

\def\rhomeson {{\ensuremath{\Prho}}\xspace}
\def\rhoz     {{\ensuremath{\rhomeson^0}}\xspace}

\def\kaon    {{\ensuremath{\PK}}\xspace}

\def\KorKbar {\kern \thebaroffset\optbar{\kern -\thebaroffset \PK}{}\xspace}

\def\Kp      {{\ensuremath{\kaon^+}}\xspace}
\def\Km      {{\ensuremath{\kaon^-}}\xspace}

\def\KS      {{\ensuremath{\kaon^0_{\mathrm{S}}}}\xspace}

\def\D       {{\ensuremath{\PD}}\xspace}

\def\DorDbar {\kern \thebaroffset\optbar{\kern -\thebaroffset \PD}\xspace}

\def\Dp      {{\ensuremath{\D^+}}\xspace}
\def\Dm      {{\ensuremath{\D^-}}\xspace}

\def\DpDm    {\ensuremath{\Dp {\kern -0.16em \Dm}}\xspace}

\def\B       {{\ensuremath{\PB}}\xspace}

\def\BorBbar {\kern \thebaroffset\optbar{\kern -\thebaroffset \PB}\xspace}

\def\Bd      {{\ensuremath{\B^0}}\xspace}

\def\BdorBdbar {\kern \thebaroffset\optbar{\kern -\thebaroffset \Bd}\xspace}

\def\Bs      {{\ensuremath{\B^0_\squark}}\xspace}

\def\BsorBsbar {\kern \thebaroffset\optbar{\kern -\thebaroffset \Bs}\xspace}

\def\Y#1S{\ensuremath{\PUpsilon{(#1S)}}\xspace}

\def\proton      {{\ensuremath{\Pp}}\xspace}

\def\LorLbar     {\kern \thebaroffset\optbar{\kern -\thebaroffset \PLambda}\xspace}

\def\AT#1     {\ensuremath{A_{\mathrm{T}}^{#1}}\xspace}           

\def\C#1      {\ensuremath{\mathcal{C}_{#1}}\xspace}                       
\def\Cp#1     {\ensuremath{\mathcal{C}_{#1}^{'}}\xspace}                    
\def\Ceff#1   {\ensuremath{\mathcal{C}_{#1}^{\mathrm{(eff)}}}\xspace}        
\def\Cpeff#1  {\ensuremath{\mathcal{C}_{#1}^{'\mathrm{(eff)}}}\xspace}       
\def\Ope#1    {\ensuremath{\mathcal{O}_{#1}}\xspace}                       
\def\Opep#1   {\ensuremath{\mathcal{O}_{#1}^{'}}\xspace}                    

\newcommand{\nospaceunit}[1]{\ensuremath{\text{#1}}}       
\newcommand{\aunit}[1]{\ensuremath{\text{\,#1}}}       

\newcommand{\tev}{\aunit{Te\kern -0.1em V}\xspace}
\newcommand{\gev}{\aunit{Ge\kern -0.1em V}\xspace}
\newcommand{\mev}{\aunit{Me\kern -0.1em V}\xspace}
\newcommand{\kev}{\aunit{ke\kern -0.1em V}\xspace}
\newcommand{\ev}{\aunit{e\kern -0.1em V}\xspace}
 
\newcommand{\mevc}{\ensuremath{\aunit{Me\kern -0.1em V\!/}c}\xspace}
\newcommand{\gevc}{\ensuremath{\aunit{Ge\kern -0.1em V\!/}c}\xspace}
\newcommand{\mevcc}{\ensuremath{\aunit{Me\kern -0.1em V\!/}c^2}\xspace}
\newcommand{\gevcc}{\ensuremath{\aunit{Ge\kern -0.1em V\!/}c^2}\xspace}

\def\mm   {\aunit{mm}\xspace}

\def\mum  {\ensuremath{\,\upmu\nospaceunit{m}}\xspace}

\def\fm   {\aunit{fm}\xspace}

\def\gsim{{~\raise.15em\hbox{$>$}\kern-.85em
          \lower.35em\hbox{$\sim$}~}\xspace}
\def\lsim{{~\raise.15em\hbox{$<$}\kern-.85em
          \lower.35em\hbox{$\sim$}~}\xspace}

\def\sqsnn {\ensuremath{\protect\sqrt{s_{\scriptscriptstyle\text{NN}}}}\xspace}
\def\pt         {\ensuremath{p_{\mathrm{T}}}\xspace}

\def\ptot       {\ensuremath{p}\xspace}

\def\evtgen     {\mbox{\textsc{EvtGen}}\xspace}

\def\geant      {\mbox{\textsc{Geant4}}\xspace}

\def\photos     {\mbox{\textsc{Photos}}\xspace}

\def\tell1  {TELL1\xspace}
\def\ukl1   {UKL1\xspace}

\newcommand{\eg}{\mbox{\itshape e.g.}\xspace}
\newcommand{\ie}{\mbox{\itshape i.e.}\xspace}

\newcommand{\lhcborcid}[1]{\href{https://orcid.org/#1}{\hspace*{0.1em}\raisebox{-0.45ex}{\includegraphics[width=1em]{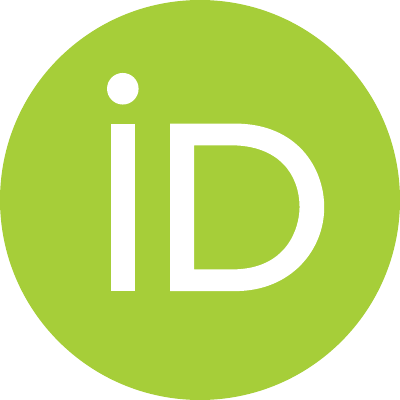}}}}

\usepackage{cite} 
\usepackage{mciteplus}

\usepackage{longtable} 

\newcommand{\myMath}[1]{\mbox{\ensuremath{#1}}}

\newcommand{\myRangeInMath}[2]{#1\text{--}#2}

\newcommand{\sigSyst}[1]{\myMath{ \sigma_{\text{syst}} (#1) }}

\newcommand{\pid}{PID\xspace}
\newcommand{\nonFemtoscopic}{nonfemtoscopic\xspace}

\newcommand{\epos}{\mbox{\textsc{EPOS}}\xspace}

\newcommand{\resonanceKS}{{\ensuremath{\KS(497)}}\xspace}
\newcommand{\resonanceRho}{{\ensuremath{\rhoz(770)}}\xspace}
\newcommand{\resonanceFZero}{{\ensuremath{f_{0}(980)}}\xspace}
\newcommand{\resonanceFTwo}{{\ensuremath{f_{2}(1270)}}\xspace}

\newcommand{\pp}{\mbox{\proton\proton}\xspace}
\newcommand{\pPb}{\mbox{\proton{}Pb}\xspace}
\newcommand{\pbP}{\mbox{Pb\proton}\xspace}

\newcommand{\pv}{\mbox{PV}\xspace}

\newcommand{\becR}{\ensuremath{R}\xspace}

\newcommand{\becLambda}{\ensuremath{\lambda}\xspace}

\newcommand{\becN}{\ensuremath{N}\xspace}  
\newcommand{\becDelta}{\ensuremath{\delta}\xspace}

\newcommand{\becREff}{\ensuremath{R_{\text{eff}}}\xspace}  
\newcommand{\becBkgWidth}{\ensuremath{\sigma_{\text{bkg}}}\xspace}
\newcommand{\becBkgAmpl}{\ensuremath{A_{\text{bkg}}}\xspace}
\newcommand{\becBkgScaling}{\ensuremath{z}\xspace}
\newcommand{\becBkgSigZero}{\ensuremath{\sigma_{0}}\xspace}
\newcommand{\becBkgSigOne}{\ensuremath{\sigma_{1}}\xspace}
\newcommand{\becBkgMultZero}{\ensuremath{N_{0}}\xspace}
\newcommand{\becBkgAmplZero}{\ensuremath{A_{0}}\xspace}
\newcommand{\becBkgAmplMultExp}{\ensuremath{n_{A}}\xspace}

\newcommand{\becQ}{\ensuremath{Q}\xspace}  

\newcommand{\corrFuncFullPlural}{correlation functions\xspace}
\newcommand{\corrFunc}{\myMath{C_2(Q)}\xspace}

\newcommand{\corrFuncGen}[2]{\myMath{C_{#1} \left( #2 \right) }\xspace}

\newcommand{\corrFuncCoulomb}{\myMath{K(Q)}\xspace}
\newcommand{\corrFuncBkg}{\myMath{\Omega(Q)}\xspace}

\newcommand{\nVelo}{\ensuremath{N_{\text{\velo}}}\xspace}
\newcommand{\nVeloFull}{\mbox{\velo-track} multiplicity\xspace}

\newcommand{\nChFull}{\mbox{charged-particle} multiplicity\xspace}

\newcommand{\bec}{\mbox{BEC}\xspace}

\newcommand{\like}{\mbox{SS}\xspace}
\newcommand{\unlike}{\mbox{OS}\xspace}

\newcommand{\zPv}{\myMath{z_{\text{\pv}}}\xspace}

\newcommand{\slopeDiffX}{\ensuremath{\Delta t_{x}}\xspace}
\newcommand{\slopeDiffY}{\ensuremath{\Delta t_{y}}\xspace} 
\newcommand{\slopeDiffXDef}{\ensuremath{\slopeDiffX = \ptot_{x_{1}} \, / \, \ptot_{z_{1}} - \ptot_{x_{2}} \, / \, \ptot_{z_{2}}}\xspace}
\newcommand{\slopeDiffYDef}{\ensuremath{\slopeDiffY = \ptot_{y_{1}} \, / \, \ptot_{z_{1}} - \ptot_{y_{2}} \, / \, \ptot_{z_{2}}}\xspace}

\newcommand{\multTypical}{\myMath{ 35 \leqslant \nVelo < 40  }\xspace}
\newcommand{\multHigh}{\myMath{ 100 \leqslant \nVelo < 115  }\xspace}

\begin{document}

\renewcommand{\thefootnote}{\fnsymbol{footnote}}
\setcounter{footnote}{1}


\begin{titlepage}
\pagenumbering{roman}

\vspace*{-1.5cm}
\centerline{\large EUROPEAN ORGANIZATION FOR NUCLEAR RESEARCH (CERN)}
\vspace*{1.5cm}
\noindent
\begin{tabular*}{\linewidth}{lc@{\extracolsep{\fill}}r@{\extracolsep{0pt}}}
\ifthenelse{\boolean{pdflatex}}
{\vspace*{-1.5cm}\mbox{\!\!\!\includegraphics[width=.14\textwidth]{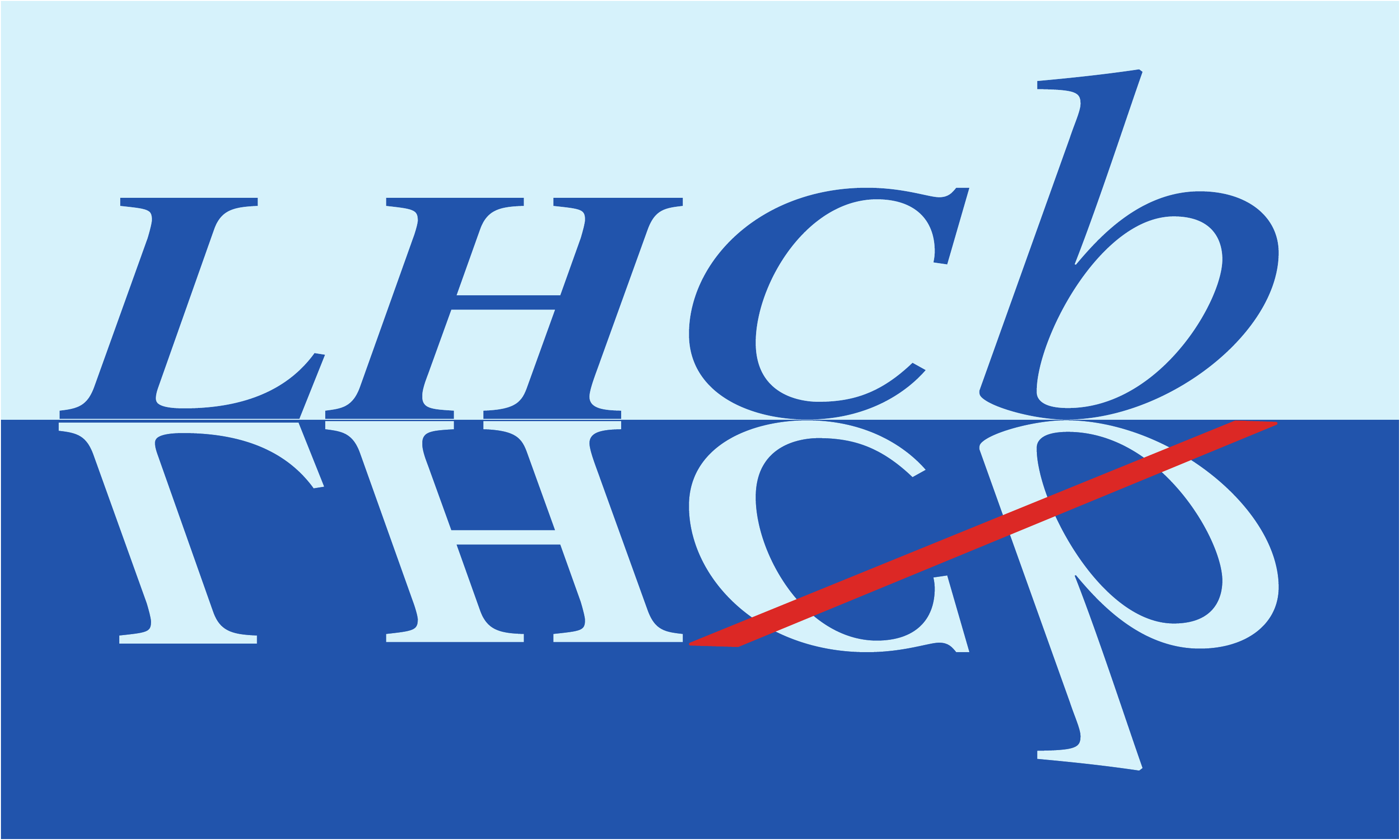}} & &}%
{\vspace*{-1.2cm}\mbox{\!\!\!\includegraphics[width=.12\textwidth]{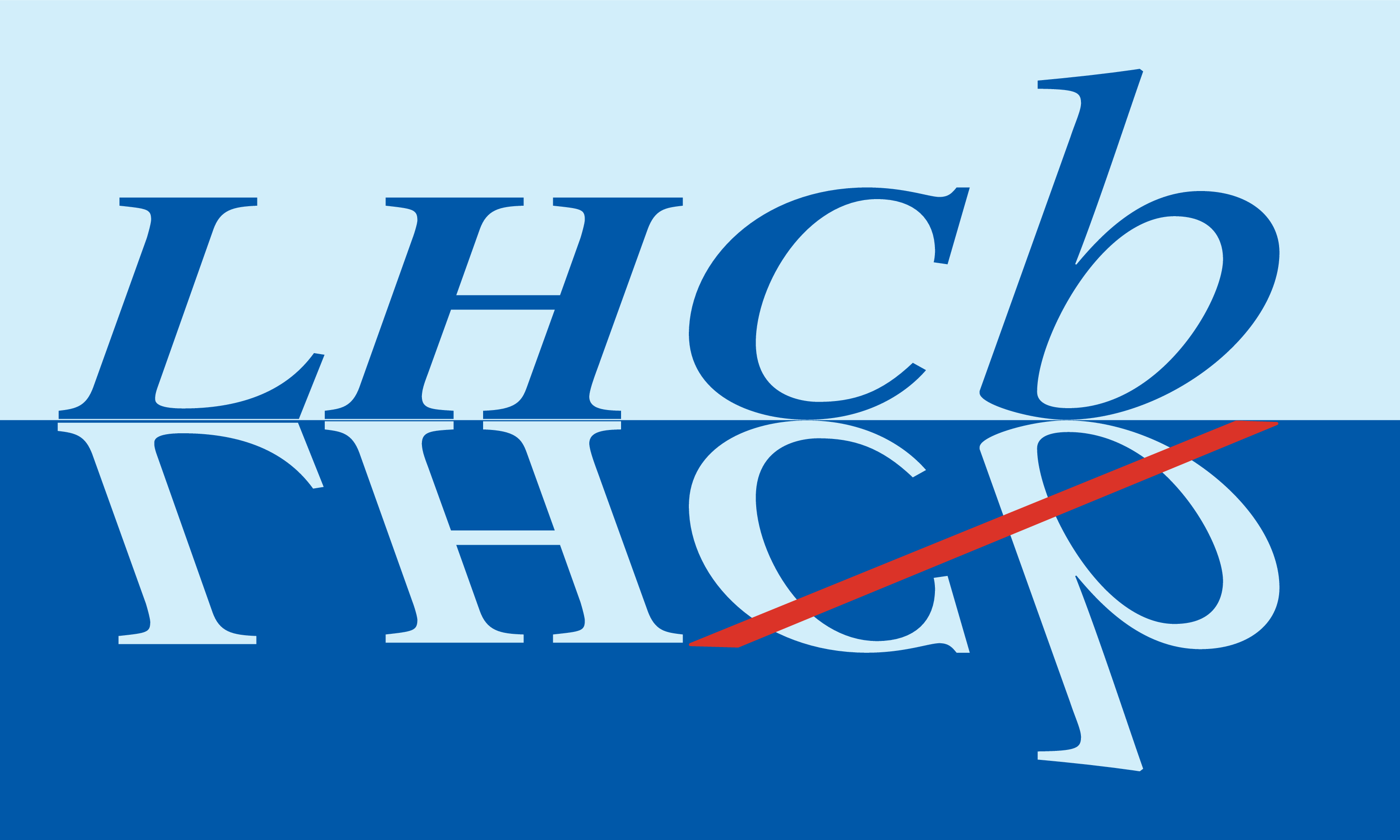}} & &}%
\\
 & & CERN-EP-2023-096 \\  
 & & LHCb-PAPER-2023-002 \\  
 & & October 9, 2023 \\ 
 & & \\
\end{tabular*}

\vspace*{4.0cm}

{\normalfont\bfseries\boldmath\huge
\begin{center}
  \papertitle 
\end{center}
}

\vspace*{2.0cm}

\begin{center}
\paperauthors
\footnote{Authors are listed at the end of this paper.}
\end{center}

\vspace{\fill}

\begin{abstract}
  \noindent

  Correlations of same-sign charged pions are analysed using proton-lead collision data collected by the \lhcb experiment at a nucleon-nucleon centre-of-mass energy of 5.02~TeV, corresponding to an integrated luminosity of 1.06~nb$^{-1}$. Bose-Einstein correlations are observed in the form of an enhancement of pair production for same-sign charged pions with a small four-momentum difference. The dependence of the correlation radius and the intercept parameter on the reconstructed charged-particle multiplicity is investigated. The measured correlation radii scale linearly with the cube root of the reconstructed charged-particle multiplicity, being compatible with predictions of hydrodynamic models on the collision system evolution. 
\end{abstract}

\vspace*{2.0cm}

\begin{center}
  Published in JHEP 09 (2023) 172
\end{center}

\vspace{\fill}

{\footnotesize 
\centerline{\copyright~\papercopyright. \href{\paperlicenceurl}{\paperlicence}.}}
\vspace*{2mm}

\end{titlepage}


\newpage
\setcounter{page}{2}
\mbox{~}
%
%
%
%


\renewcommand{\thefootnote}{\arabic{footnote}}
\setcounter{footnote}{0}

\cleardoublepage


\pagestyle{plain} 
\setcounter{page}{1}
\pagenumbering{arabic}


\section{Introduction}
\label{sec:Introduction}

Multiparticle production in the process of hadronization has been investigated for six decades but its nature is still not fully understood. The Hanbury Brown-Twiss (HBT) intensity interferometry~\cite{HanburyBrown:1954,Brown:1956,HanburyBrown:1956} is the main tool to study the space-time properties of the hadron emission volume. In the case of joint production of identical bosons the HBT interference effect results in Bose-Einstein Correlations (BEC), while in the case of fermions it is referred to as Fermi-Dirac Correlations (FDC). The  correlations measure the scales that are referred to as lengths of homogeneity~\cite{Makhlin:1987,Sinyukov:1994}, which correspond to a limited region of the particle-emitting source surface. Measurement of  correlations of identical particles can provide insight into the evolution of the hadron source. In particular, small systems, such as those produced in proton-ion (pA) collisions, are interesting because their lifetimes are significantly shorter than those in heavy-ion (AA) collisions, thus providing a better probe of the early system dynamics and the initial geometry.

Since the first observation of quantum interference effects in identically charged pions produced in proton-antiproton collisions~\cite{Goldhaber:1959}, such effects have been investigated by many different experiments, \eg at the Berkeley Bevalac~\cite{desmond}, AGS~\cite{Ahle:2002}, ISR~\cite{Akesson:1983,Akesson:1985,Akesson:1986}, SPS~\cite{Albajar:1989,Boggild:1994,Adamova:2002}, LEP~\cite{Decamp:1991,Buskulic:1994,Barate:1999,Heister:2003,Abreu:1992,Abreu:1996,Abreu:1999,Acciarri:1999,Achard:2001,Achard:2011,Abbiendi:2000,Akers:1995,Alexander:1996,Alexander:1996a,Abbiendi:2000a,Abbiendi:2003}, RHIC~\cite{Adler:2001,Adams:2003,Adams:2003a,Adams:2004,Abelev:2009,Adamczyk:2013,Adamczyk:2014,Adler:2004,Aggarwal:2010,Afanasiev:2007,Adare:2006,Adare:2017,Back:2004},  \lhc~\cite{Khachatryan:2010,Khachatryan:2011,Sirunyan:2019,Aamodt:2010,Abelev:2012,Abelev:2012a,Acharya:2018,Aad:2015,LHCb-PAPER-2017-025,Aamodt:2011,Abelev:2014,Adam:2015,Sirunyan:2017,Acharya:2019,Acharya:2019a,Aaboud:2017,Adam:2015a,Acharya:2017,Aamodt:2011a} and others~\cite{Alexopoulos:1992,Adloff:1997}. The sizes of the studied systems vary tremendously, from \epem collisions at LEP to AuAu collisions at RHIC and PbPb beams at \lhc. At the \lhc alone, the HBT effect has been investigated in proton-proton (\pp)~\cite{Khachatryan:2010,Khachatryan:2011,Sirunyan:2019,Aamodt:2010,Abelev:2012,Abelev:2012a,Acharya:2018,Aad:2015,LHCb-PAPER-2017-025,Aamodt:2011,Abelev:2014,Adam:2015,Sirunyan:2017,Acharya:2019}, proton-lead (\pPb)~\cite{Abelev:2014,Adam:2015,Sirunyan:2017,Acharya:2019,Acharya:2019a,Aaboud:2017,Adam:2015a} and lead-lead (PbPb)~\cite{Aamodt:2011,Abelev:2014,Adam:2015,Sirunyan:2017,Adam:2015a,Acharya:2017,Aamodt:2011a} systems.

In this paper, the first study of the BEC effect in \pPb and \pbP collisions in the forward rapidity region is presented. The \lhcb detector has the potential to measure quantum interference effects in the forward region, and therefore to provide complementary results to those from the other LHC experiments in the central rapidity region. This allows to study the dependence of the quantum interference effects upon various observables, and it provides insight into the particle production process in the forward direction, useful for the development of theoretical models.

\section{Analysis method}
The BEC or FDC effects are the result of the quantum statistics, caused by the symmetrization (antisymmetrization) of the wave function describing a system of bosons (fermions). Such  correlations are examined by measuring a two-particle correlation function, defined as the ratio of the inclusive density distribution for two particles and the so-called reference density. The latter is a two-particle density distribution that approximates the distribution without the BEC or FDC effects. The present study is based on the assumption of static, spherically-symmetric sources that can be characterized by univariate distributions. This class of sources is commonly used in HBT analyses, since the measured correlation radii in this case can be interpreted as the effective size of the particle-emitting source at the kinetic freeze-out~\cite{Csorgo:1999,Kittel:2001}.
 
\subsection{Correlation function}
In order to investigate the space-time evolution of the hadronization source, the correlation function is commonly studied using the Lorentz-invariant variable $Q$~\cite{Baym:1997}, which is related to the difference in the four-momenta $q_{1}$ and $q_{2}$ of two indistinguishable particles of rest mass $m$,
\begin{equation}
	\label{eq:theory:defQ}
	\becQ \equiv \sqrt{ - \left( q_{1} - q_{2} \right)^2 } = \sqrt{M^{2} - 4 m^{2}}\ .
\end{equation}
This gives a measure of the phase-space separation of the two-particle system of invariant mass $M$. A two-particle correlation function $C_{2}$ is constructed as the ratio of the $Q$ distributions for signal and reference pairs
\begin{equation}
\label{eq:corrFunc:def}
\corrFunc= \left( \frac{N^{\rm ref}}{N^{\rm sig}} \right) \left( \frac{ {\rm d} N^{\rm sig}(\becQ) \, / \, {\rm d} \becQ }{ {\rm d} N^{\rm ref}(\becQ) \, / \, {\rm d} \becQ } \right) \ ,
\end{equation}
where \myMath{N^{\rm sig}} and \myMath{N^{\rm ref}} correspond to the number of signal and reference pairs, respectively, obtained from an integral of the relevant $Q$ distributions. Signal pairs are formed from selected  same-sign (SS) charged particles that originate from the same collision vertex where the Bose-Einstein correlations are expected. The reference pairs are pairs of pions which reproduce as closely as possible the kinematics and various effects present in the signal, except for the BEC effect. The correlation function is constructed as a ratio to cancel the effects related to the detection efficiency.

There are several methods to obtain a reference sample. It can be constructed using experimental data, or with simulated events incorporating the detector interactions. In this study, a data-driven reference sample is constructed by collecting pairs of particles originating from different collision events (a so-called event-mixing method), where the BEC effect cannot be present. The reference pairs are selected in the similar way as the signal ones to ensure that the signal kinematic distributions are reproduced as closely as possible.  Additional requirements are imposed to combine particles originating from different events with similar properties and to further improve the agreement between the signal and reference samples. Particles in the reference pairs are required to originate from primary vertices with a comparable multiplicity of tracks reconstructed in the vertex detector. Event-mixing candidates for the current event are provided by creating a pool of selected particles from ten different events and splitting them into intervals of multiplicity and the coordinate of the primary vertex (PV) along the beam direction $z_{PV}$. Then, for each signal pair in a PV with a given (multiplicity, $z_{PV}$) interval, a random particle is chosen from the relevant interval in the pool to create a reference pair with the particle in the signal pair. Candidate particles to create pairs are grouped into multiplicity bins of width of three. The distance between the two PVs associated with the two particles in the same reference pair must be smaller than 10~mm in the $z$ direction to provide consistent detector acceptance effects for both particles.

The two-particle correlation function associated with a general class of particle sources can be described by the so-called symmetric L\'evy-stable distributions~\cite{Csorgo:2003}. In the case of static, univariate sources, the L\'evy-type correlation function is expressed as
\begin{equation}
	\label{eq:theory:corrFunc:levy}
	\corrFuncGen{2,\text{\bec}}{\becQ} = 1 + e^{ - \left| \becR \becQ \right|^{{\alpha}_{ \text{\miniscule \it L} }}} \ ,
\end{equation}
where $R$ denotes the correlation radius, and $\alpha_{L}$ is a parameter that can take values in the range $0 < \alpha_{L} < 2$ and is referred to as a L\'evy index of stability. Frequently, to enable comparison of the correlation parameters between experiments and between different collision systems, including a measurement by LHCb in \pp collisions~\cite{LHCb-PAPER-2017-025}, $\alpha_{L}$ is fixed to one, leading to the simplified expression:
\begin{equation}
	\label{eq:theory:corrFunc:exp}
	\corrFuncGen{2,\text{\bec}}{\becQ} = 1 + e^{ - \left| \becR \becQ \right|} \ .
\end{equation}
This parametrization enables the measured correlation radius to be interpreted as the effective size of the particle-emitting source.

\subsection{Final state interactions and nonfemtoscopic effects}
Final state interactions (FSI) resulting from the strong and electromagnetic forces can affect the observed two-particle correlations. The effects of the strong interaction in the case of pions is relatively small~\cite{Bowler:1991} and is usually neglected in BEC studies. The most notable effect is Coulomb repulsion related to the same-sign electric charge of the studied particles, especially in the low-$Q$ region. A general expression for the Coulomb interaction term for point-like sources~\cite{Bowler:1991,Sinyukov:1998,Csanad:2019}, $K(Q)$, is equivalent to the so-called Gamov factor~\cite{Pratt:1986} for same-sign (SS) and opposite-sign (OS) pairs:
\begin{equation}
    \label{eq:coulomb:gamovFactor}
    K_{\rm Gamov}^{\text{\like}}(\zeta) = \frac{ 2 \pi \zeta }{ e^{ 2 \pi \zeta } - 1 } \,\, , \,\, K_{\rm Gamov}^{\text{\unlike}}(\zeta) = \frac{ 2 \pi \zeta }{ 1 - e^{ - 2 \pi \zeta }} \ ,
\end{equation}
where \myMath{\zeta =  \alpha m \, / \, \becQ }, \myMath{\alpha} is the~\mbox{fine-structure} constant and \myMath{m} is the particle's rest mass. For SS particles, a repulsive interaction leads to a decrease in the correlation function, which is most prominent for low $Q$-values. In the case of OS pairs, this effect is reversed and an enhancement is observed. The OS sample is useful to parametrize the background related to the cluster contribution, as explained later.

In the present analysis a parametrization developed by the \cms experiment~\cite{Sirunyan:2017}, valid for the L\'evy-type sources with $\alpha_{L}$ equal to unity, is used to account for the final-state Coulomb interactions between the particles in the SS and OS pairs
\begin{equation}
    \label{eq:coulomb:cms}
    \corrFuncCoulomb = K_{\rm Gamov}(Q) \, \left( 1 + \frac{\alpha \pi m \becREff}{1.26 + \becQ \becREff} \right) \ ,
\end{equation}
where \becREff corresponds to the~effective size of the particle-emitting source and is provided in femtometres. The additional term with \becREff represents a correction to the Gamov factor that enables a more precise characterization of the Coulomb interaction for extended sources.

The correlation function shape is distorted by the presence of various nonfemtoscopic\footnote{Femtoscopic effects are those observed at the Fermi scale.} effects. There is no strict, theory-motivated description of such contributions, and different strategies can be applied to take them into account in the analysis (see Sec.~5). Long-range correlations, being one of nonfemtoscopic effects related mostly to the energy-momentum conservation, are present in the full $Q$ range, but are most prominent at the high-$Q$ values~(\myMath{\sim \becQ > 1 \gev})\footnote{If not indicated otherwise natural units with $c$ = 1 are used.}, far from the BEC-signal region. Although different parametrizations can be employed, a simple term linear in $Q$ is usually optimal to characterize this contribution and is therefore commonly used~\cite{Kittel:2001zw,Alexander:2003ug}.

Cluster contribution~\cite{Sirunyan:2017} is another prominent component of the nonfemtoscopic background, related to the effects of particles emitted inside low-momentum mini-jets and multibody decays of resonances. It is difficult to correct for the long-range correlations, as these are  present dominantly in the range $Q < 0.5$--1.0~GeV that overlaps with the BEC signal. Constructing a correlation function for OS pairs can be particularly useful in the background studies, since similar effects can be expected for both the SS and OS pairs. Special care must be taken when investigating OS pairs, due to structures related to two-body decays of resonances, arising in the correlation function.

Nonfemtoscopic background effects in the present analysis (in particular the cluster contribution) are studied and parametrized using the OS correlation functions. A cluster subtraction (CS) method, which was developed by the CMS experiment~\cite{Sirunyan:2017,Sirunyan:2019}, is employed for this purpose. This technique represents a fully data-driven approach. In the CS method, the shape parameters of the chosen function for the background description (e.g. the width of a Gaussian distribution) are determined from the OS fits. The background shape parameters in the SS fits are fixed to the values determined by studying the OS pairs, with an additional scaling parameter that is introduced to account for the different amplitudes of the cluster contribution in the SS and OS correlation functions.

\subsection{Fitting method}
The correlation function, including electromagnetic effects and the nonfemoscopic background, is parameterized using the Bowler–Sinyukov formalism~\cite{Bowler:1991,Sinyukov:1998}, as 
\begin{equation}
	\label{eq:pPb:corrFunc:like}
  \corrFunc = \becN \left[ 1 - \becLambda + \becLambda \corrFuncCoulomb \times \left( 1 + e^{ - \left| \becR \becQ \right| } \right) \right] \times \corrFuncBkg \ ,
\end{equation}
where \becN~is a~normalization factor and \myMath{\corrFuncBkg} is a~general term for the~\nonFemtoscopic background contribution, as described later. The intercept parameter,   $\lambda$, corresponds to the extrapolated value of the correlation function at $Q = 0$~GeV~\cite{Csorgo:1999}. This observable can be interpreted within the core-halo model~\cite{Csorgo:1994in}, which assumes that the particle emission can take place in a central core or in an extended halo originating from long-lived resonance decays.

Contents of the bins in both the signal and reference $Q$-variable histograms are Poisson-distributed, and hence a negative log-likelihood fit method is preferable for the BEC studies~\cite{Ahle:2002,Aaboud:2017}. In this approach, the following expression is minimized in the fitting procedure:
\begin{equation}
    \label{eq:logLikelihood}
    -2 \ln L = 2 \sum_{i}  \left\{ A_i \ln \left[ \frac{ \left( 1 + C_{2i} \right) A_i }{ C_{2i} \left( A_i + B_i + 2 \right) } \right] + \left( B_i + 2 \right) \ln \left[ \frac{ \left( 1 + C_{2i} \right) \left( B_i + 2 \right) }{ A_i + B_i + 2 } \right] \right\}  \ ,
\end{equation}
where \myMath{A_i} and \myMath{B_i} are the~bin contents of the~signal and reference \myMath{\becQ} histograms and \myMath{C_{2i}} corresponds to the fitted value of the correlation function at the $Q$-bin centre.

\section{Detector and dataset}
The LHCb detector~\cite{LHCb-DP-2008-001,LHCb-DP-2014-002} is a single-arm forward spectrometer covering the pseudorapidity range $2 < \eta < 5$, designed for the study of particles containing $b$ or $c$ quarks. The detector includes a high-precision tracking system consisting of a silicon-strip vertex detector (VELO) surrounding the  interaction region~\cite{LHCb-DP-2014-001}, a large-area silicon-strip detector located upstream of a dipole magnet with a bending power of about $4{\rm\,Tm}$, and three stations of silicon-strip detectors and straw drift tubes~\cite{LHCb-DP-2013-003} placed downstream of the magnet. The tracking system provides a measurement of the momentum, \ptot, of charged particles with a relative uncertainty that varies from 0.5\% at low momentum to 1.0\% at 200~GeV. The minimum distance of a track to a primary collision vertex (PV), the impact parameter, is measured with a resolution of (15 + $29 /$ \pt) \mum, where \pt is the component of the momentum transverse to the beam, in \gev. Different types of charged hadrons are distinguished using information from two ring-imaging Cherenkov detectors~\cite{LHCb-DP-2012-003}. Photons, electrons and hadrons are identified by a calorimeter system consisting of scintillating-pad and preshower detectors, an electromagnetic and a hadronic calorimeter. Muons are identified by a system composed of alternating layers of iron and multiwire proportional chambers~\cite{LHCb-DP-2012-002}. The trigger~\cite{LHCb-DP-2012-004} consists of a hardware stage, based on information from the calorimeter and muon systems, followed by a software stage, which applies a full event reconstruction.

In the present analysis, a dataset of minimum-bias triggered events collected in \pPb collisions recorded in 2013 at a nucleon-nucleon centre-of-mass energy \sqsnn = 5.02~\tev is used, with 4 and 1.58~\tev beam energies, respectively. Two collision modes were used in this data-taking period with the beam directions reversed, which permits the study of \pPb collisions both in the forward and backward rapidity regions. The recorded \pPb and \pbP samples correspond to integrated luminosities of 1.06~nb$^{-1}$ and 0.52~nb$^{-1}$, respectively. As only a fraction of $\sim 10^{-5}$ of the collisions corresponds to multiple interactions, a dedicated selection requirement is applied to accept only the events with a single primary vertex. The data samples available in the current study after selection described in Sec.~\ref{evtSel} corresponds to $\sim 6.3 \times 10^{7}$ ($5.7 \times 10^{7}$) events for the \pPb (\pbP) collisions.

Simulation samples corresponding to the 2013 \pPb data-taking conditions are produced using the \epos-LHC~\cite{Pierog:2013} generator, with a specific LHCb configuration~\cite{Belyaev:2011}. Decays of hadronic particles are described by \evtgen~\cite{Lange:2001}, in which final-state radiation is generated using \photos~\cite{Golonka:2006}. The interaction of the generated particles with the detector and its response are implemented using the \geant toolkit~\cite{Allison:2006,Agostinelli:2003}, as described in~\cite{Clemencic:2011}. The BEC effect is not activated in the simulation. Each of the \pPb and \pbP simulated dataset contains $\sim 1.2 \times 10^{7}$ events after selection, with the number of interactions per bunch crossing fixed to unity. The simulation samples are used mainly for the event-selection optimization, while the background modelling is performed with a purely data-driven approach.

\section{Event selection}
\label{evtSel}
The data samples are divided into bins of the multiplicity of tracks reconstructed in the vertex detector (VELO tracks) assigned to a PV ($N_{\rm VELO}$), which is used as a proxy variable to describe the total charged-particle multiplicity produced at the PV. The division is optimized to obtain a high number of bins with enough entries to perform the measurement (see appendix~\ref{app:mult}). The chosen binning scheme is presented in Table~\ref{tab:pPb:multiplicityBins}, together with approximate fractions of the respective sample corresponding to the given bin.

\begin{table}[tp]
	\caption{
		\small 
		Ranges of the~\nVeloFull bins in the~\pPb and \pbP datasets. An~approximate fraction of the~relevant~data sample corresponding to the~given bin is also indicated.
	}
	\begin{center}
		\vspace*{-0.5cm}
		\begin{tabular}{cccc}
    \hline
    & & \multicolumn{2}{c}{Sample fraction [\myMath{\%}]} \\
    bin\# & \nVelo & \pPb & \pbP \\
    \hline
    1       & \myMath{\myRangeInMath{5}{9}}     & \myMath{ < 2 }        & \myMath{ < 2 }    \\
    2       & \myMath{\myRangeInMath{10}{14}}     & \myMath{ 2 }          & \myMath{ 2 }      \\
    3       & \myMath{\myRangeInMath{15}{19}}     & \myMath{ 4 }          & \myMath{ 2 }      \\
    4       & \myMath{\myRangeInMath{20}{24}}     & \myMath{ 7 }          & \myMath{ 3 }      \\
    5       & \myMath{\myRangeInMath{25}{29}}     & \myMath{ 10 }         & \myMath{ 4 }      \\
    6       & \myMath{\myRangeInMath{30}{34}}     & \myMath{ 13 }         & \myMath{ 5 }      \\
    7       & \myMath{\myRangeInMath{35}{39}}     & \myMath{ 14 }         & \myMath{ 6 }      \\
    8       & \myMath{\myRangeInMath{40}{44}}     & \myMath{ 10 }         & \myMath{ 5 }      \\
    9       & \myMath{\myRangeInMath{45}{49}}     & \myMath{ 10 }         & \myMath{ 6 }      \\
    10      & \myMath{\myRangeInMath{50}{54}}     & \myMath{ 8 }          & \myMath{ 6 }      \\
    11      & \myMath{\myRangeInMath{55}{59}}     & \myMath{ 7 }          & \myMath{ 7 }      \\
    12      & \myMath{\myRangeInMath{60}{64}}     & \myMath{ 5 }          & \myMath{ 6 }      \\
    13      & \myMath{\myRangeInMath{65}{79}}     & \myMath{ 6 }          & \myMath{ 15 }     \\
    14      & \myMath{\myRangeInMath{80}{89}}     & \myMath{ \text{--} }  & \myMath{ 7 }      \\
    15      & \myMath{\myRangeInMath{90}{99}}    & \myMath{ \text{--} }  & \myMath{ 7 }      \\    
    16      & \myMath{\myRangeInMath{100}{114}}   & \myMath{ \text{--} }  & \myMath{ 6 }      \\    
    17      & \myMath{\myRangeInMath{115}{139}}   & \myMath{ \text{--} }  & \myMath{ 7 }      \\    
    18      & \myMath{\myRangeInMath{140}{179}}   & \myMath{ \text{--} }  & \myMath{ 4 }      \\    
    \hline
\end{tabular} 
	\end{center}
	\label{tab:pPb:multiplicityBins}
\end{table}

Event selections are first applied to single-pion candidates. All pion candidates must have reconstructed track segments in both the VELO detector and tracking stations downstream of the magnet, have no matching tracks in the muon stations, and be in the pseudorapidity range $2 < \eta < 5$. Each track must have a good track-fit quality and \mbox{\pt$> 0.1$~\gev}. To suppress the contribution from secondary pions (those not associated to a PV), the impact parameter is required to be less than 0.4~mm. Furthermore, the PV is required to be located within \myMath{ -160 < \zPv < 60\mm }.

The particle identification (PID) is based on the output of a neural network employing subdetector information that quantifies the probability ProbNN for a particle to be of a certain kind~\cite{LHCb-DP-2012-003}. The simulated quantities are corrected using PID calibration samples in data~\cite{LHCb-PUB-2016-021}. Effects of PID correlations between particles are considered. It is important to ensure a sample with high purity, but a strict requirement on ProbNN(\pion) variable may also strongly affect the signal region of the correlation function, by suppressing low-momentum pions that contribute to the BEC effect. The nominal requirement of ProbNN(\pion) $>$ 0.65 is imposed to make this analysis consistent with the previous analysis for \pp collisions~\cite{LHCb-PAPER-2017-025}. Varying this requirement from 0.5 to 0.8 shows no significant changes in the measured correlation function. 

Contamination from incorrectly reconstructed particles can influence the measured Bose-Einstein correlations effect. Cloned tracks, being multiple tracks reconstructed from hits that were deposited by a single charged particle, are especially detrimental as they are present mostly in the low-$Q$ region~(\myMath{ \becQ < 1.0 \gev}), where the BEC signal is expected, appearing as a pair of almost identical, seemingly correlated, particles. To control this effect, the slopes of the track are studied. The cloned tracks usually share a very similar trajectory, hence the differences in the relevant slopes in a particle pair~(\slopeDiffXDef and~\slopeDiffYDef) tend to be small. A requirement is imposed to limit this contribution, \ie if both $|$\slopeDiffX$|$ and $|$\slopeDiffY$|$ values are smaller than $0.3 \times 10^{-3}$, then the pair is discarded. After applying these requirements, the effect of the clone particles is found to be negligible in the region $Q > 0.05$~GeV. In order to further reduce the contamination from cloned tracks and fake tracks (which do not correspond to any particle trajectory, but are reconstructed from a number of unrelated hits), in the case where the tracks share all the same hits deposited in the VELO subdetector, only the track with the best $\chi^{2}$/ndf is retained.

The study of the correlations is limited to the $Q$ range from 0.05 to 2.0~GeV. In the region with very low $Q$ ($< 0.05$~GeV) the separation in the momentum between the particles is poor and the discrepancy between simulation and data grows as $Q$ vanishes. Furthermore, investigations using simulation indicate that there is a significant fraction of pion pairs containing fake tracks and cloned tracks in the region $Q < 0.05$~GeV for all multiplicity bins.

\section{Fitting correlation functions}
Correlation functions for both the SS and OS pairs are constructed for $Q$ values between 0.05-2.0~\gev with a bin width of 0.005~\gev. This particular choice enhances consistency with the study performed for \pp collisions~\cite{LHCb-PAPER-2017-025} and allows a direct comparison of the results of the two analyses. The correlation function for the SS pairs is studied by fitting the $Q$ spectrum using Eq.~\ref{eq:pPb:corrFunc:like}. The effective radius \mbox{\becREff} in Eq.~\ref{eq:coulomb:cms} is set to 2~\fm, based on the expected correlation radii. The description of nonfemtoscopic background effects is found using the correlation function for the OS pairs. The resulting contribution is then scaled and fixed in the final fits to the SS correlation functions, as explained in detail below.

The presence of structures related to intermediate states, such as \resonanceRho, \resonanceKS, \resonanceFZero, \resonanceFTwo, in the OS correlation functions degrades the quality of the fit from which the nonfemtoscopic background parameters are determined. Therefore, the affected regions are removed from the fit to the correlation function. The boundaries of the removed regions are optimized to provide a good quality of fits to the correlation function for OS pairs, and the choice of particular boundary values is accounted for in the study of the systematic uncertainties. It is worth noting that the impact of resonances is most prominent in the bins with low $N_{\rm VELO}$ values. The observed structures quickly diminish with increasing multiplicity due to a prevailing contribution from pairs of unrelated particles.

\begin{figure}[tbp]
    \begin{center}
      \includegraphics[width=0.49\linewidth]{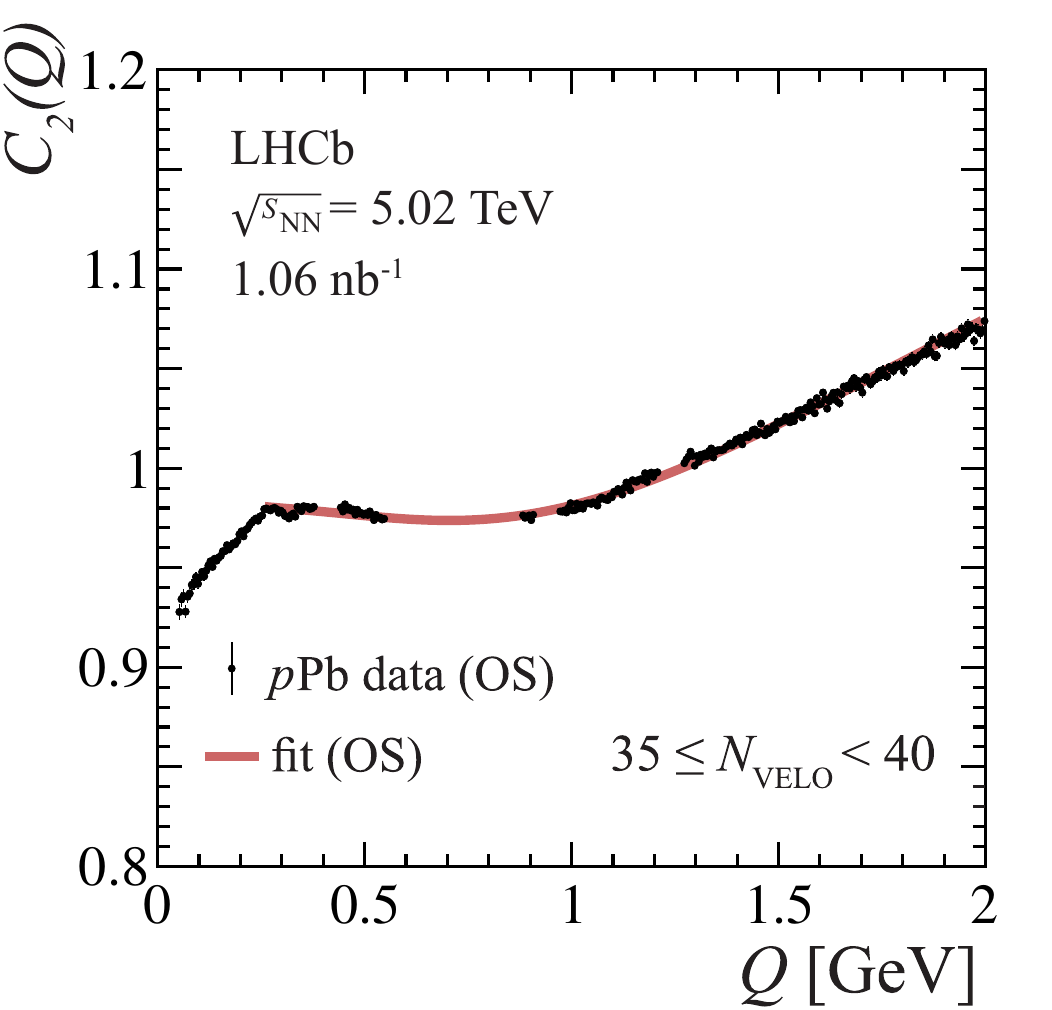}
      \includegraphics[width=0.49\linewidth]{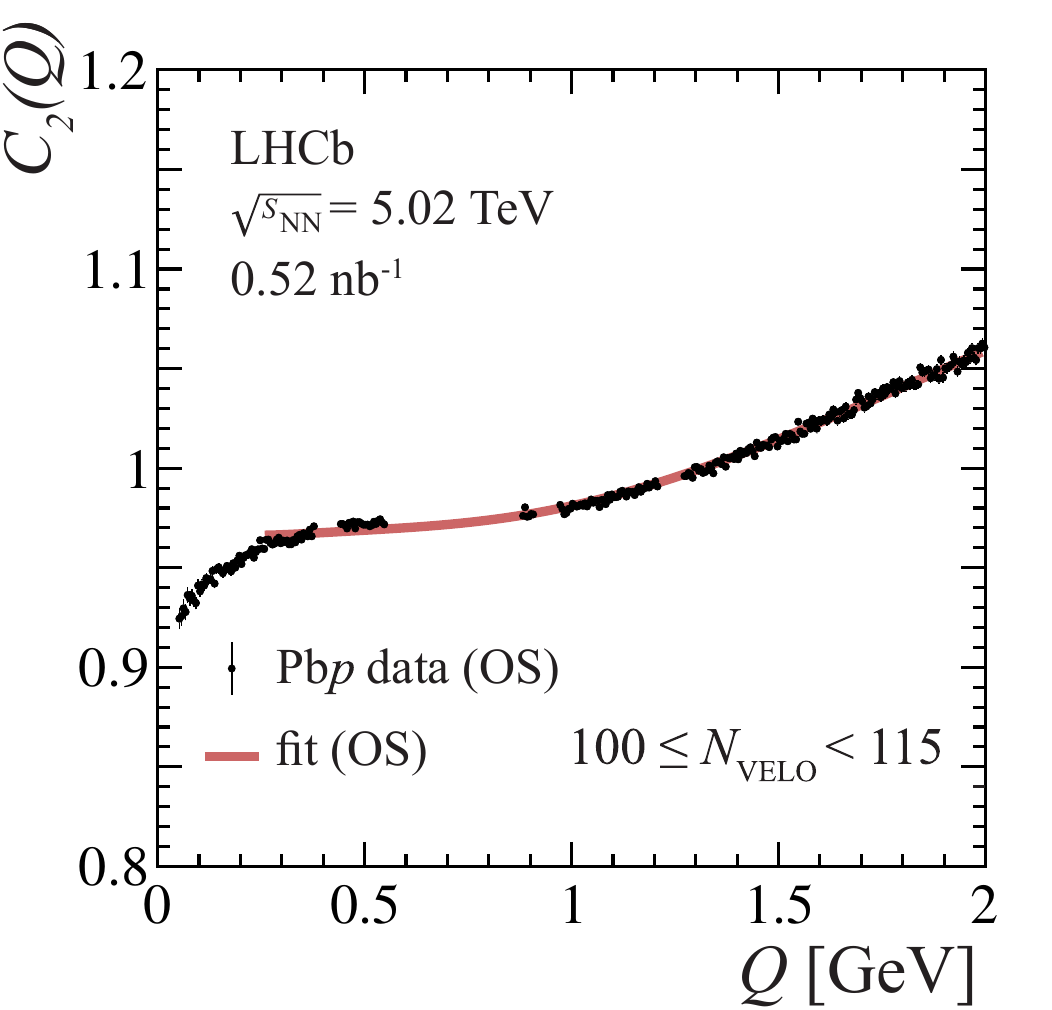}
      \vspace*{-0.5cm}
    \end{center}
    \caption{
      \small 
      Example of the~\unlike \corrFuncFullPlural in individual \nVelo bins together with the~global fit using Eq.~\ref{eq:pPb:corrFunc:bkg} to parametrize the~cluster contribution. The~results are shown for (left)~a~\mbox{moderate-multiplicity} region~(\multTypical) of the~\pPb and (right)~\mbox{a high-multiplicity} regime~(\multHigh) of the~\pbP dataset. Only statistical uncertainties are shown.
    }	
    \label{fig:pPb:results:fits:unlike}
\end{figure}

As it has been already mentioned, the OS correlation functions contain similar effects as the SS ones (apart from the BEC signal) and can be used to investigate the nonfemtoscopic background contribution. A satisfactory description of the data is found using a Gaussian parametrization for the cluster contribution~\cite{Sirunyan:2017} 
\begin{equation}
	\label{eq:pPb:corrFunc:bkg}
  \corrFuncBkg = \left( 1 + \becDelta \becQ \right) \times \left[ 1 + \becBkgScaling \frac{\becBkgAmpl}{\becBkgWidth \sqrt{2 \pi}} \exp \left( -\frac{\becQ^{2}}{2 \becBkgWidth^{2}} \right) \right]\ ,
\end{equation}
where the term linear in \becDelta corresponds to the long-range correlations and the $z$ parameter (fixed to unity in the OS fits) is a factor used for the background scaling between OS and SS pairs. The width \becBkgWidth and amplitude \becBkgAmpl are multiplicity-dependent values that characterize the cluster contribution and are parametrized as~\cite{Sirunyan:2017,Sirunyan:2019}
\begin{equation}
	\label{eq:pPb:corrFunc:bkg:width}
  \becBkgWidth \left( \nVelo \right) = \becBkgSigZero + \becBkgSigOne \exp \left( -\frac{ \nVelo }{ \becBkgMultZero } \right) \ ,
\end{equation}
\begin{equation}
	\label{eq:pPb:corrFunc:bkg:amplitude}
  \becBkgAmpl \left( \nVelo \right) = \frac{ \becBkgAmplZero }{ {\left(\nVelo\right)}^{ \becBkgAmplMultExp } } \ .
\end{equation}
The fits to the correlation functions for OS pairs are performed simultaneously in all  multiplicity bins available in the given sample (separately for the \pPb and \pbP datasets). In this procedure, the parameters from Eqs.~\ref{eq:pPb:corrFunc:bkg:width} and~\ref{eq:pPb:corrFunc:bkg:amplitude} are common for all bins, while the $N$ and $\delta$ values (Eqs.~\ref{eq:pPb:corrFunc:like} and~\ref{eq:pPb:corrFunc:bkg} respectively) are left free for each correlation function. A negative log-likelihood function (see Eq.~\ref{eq:logLikelihood}) is constructed for all the $N_{\rm VELO}$ bins in the given dataset and minimized globally to obtain the best description of the data. The~lower $Q$ fit range for the~\unlike pairs is limited with respect to the~\like ones, due to a~significant contribution of multibody resonance decays in the~very low \becQ region~\cite{Aaboud:2017,Sirunyan:2019}. The~global fits are performed for \becQ $>$ 0.25~\gev. Also, as it is found that the~best stability of the~global fits is obtained with a~fixed value of the~\becBkgMultZero parameter from Eq.~\ref{eq:pPb:corrFunc:bkg:width}, its value is set to 15 based on the fit results for the entire \pPb dataset obtained with this parameter left free. Results of the~global fits to the~\unlike \corrFuncFullPlural for the~\pPb and \pbP data are shown in Fig.~\ref{fig:pPb:results:fits:unlike} and summarized in Table~\ref{tab:pPb:results:fits:unlike}. The quality of the fits is evaluated through the normalized Baker-Cousin likelihood ratio~\cite{Baker} corresponding to the~final value of the~function minimized in the~fitting procedure~(see Eq.~\ref{eq:logLikelihood}), divided by the~number of degrees of freedom in the~fit ($\sim$2 for both \pPb and \pbP). It is worth noting that the fit quality in BEC studies is not expected to be perfect. Due to the ad hoc descriptions of the unknown nonfemtoscopic background contribution as well as the compromise between the fit quality and interpretability of the measured correlation parameters, the obtained $\chi^{2}$/ndf values are often larger than unity.

\begin{table}[tp]
    \caption{
        \small 
        Results of the~global fits to the~\unlike \corrFuncFullPlural using Eq.~\ref{eq:pPb:corrFunc:bkg} to parametrize the~cluster contribution in the \pPb and \pbP data. 
    }
    \begin{center}
        \vspace*{-0.5cm}
        \small
        \begin{tabular}{cccccc}
    \hline
    Dataset & \becBkgAmplZero [GeV] & \becBkgAmplMultExp & \becBkgSigZero [GeV] & \becBkgSigOne [GeV] \\
    \hline
    \pPb & \myMath{ 2.838 \pm 0.109 } & \myMath{ 0.8438 \pm 0.0111 } & \myMath{ 0.4799 \pm 0.0018 } & \myMath{ 0.1744 \pm 0.0060 } \\
    \pbP & \myMath{ 1.107 \pm 0.022 } & \myMath{ 0.5036 \pm 0.0049 } & \myMath{ 0.5613 \pm 0.0013 } & \myMath{ 0.0 \pm 10^{-3} } \\
    \hline
\end{tabular}
    \end{center}
    \label{tab:pPb:results:fits:unlike}
\end{table}

The cluster contribution is expected to be larger for the OS pairs than for the SS ones due to the charge conservation in processes contributing to the cluster formation. For this reason, the amplitude of the cluster contribution for the SS correlation functions is multiplied by the scaling factor $z$ (see Eq.~\ref{eq:pPb:corrFunc:bkg}). To obtain a uniform background scaling across the $N_{\rm VELO}$, this value is parametrized using a theoretically motivated form based on the ratio of SS and OS pair combinatorics~\cite{Sirunyan:2017}
\begin{equation}
	\label{eq:pPb:corrFunc:bkgScaling}
  \becBkgScaling{(\nVelo)} = \frac{ a \nVelo + b }{ 1 + a \nVelo + b  } \ ,
\end{equation}
where \myMath{a} and \myMath{b} are parameters that vary freely in the fit. Results of fits using Eq.~\ref{eq:pPb:corrFunc:bkgScaling} to determine the parametrization of the background scaling between the correlation functions for OS and SS pairs give $a = 0.044 \pm 0.004$ ($0.075 \pm 0.007$) and $b = 1.86 \pm 0.12$ ($3.12 \pm 0.27$) for \pPb (\pbP).

\section{Systematic uncertainties}
Several sources of the systematic uncertainties are studied. The values determined for each of the sources are summarized in Table~\ref{tab:pPb:systematics}, where each input is assessed by taking the difference between the refitted correlation parameters and the baseline results, excluding the sources that proved to be negligible. The general approach to determine the systematic uncertainty is to repeat the analysis procedure with appropriate modifications introduced to evaluate the contribution in question. Some $N_{\rm VELO}$ bins display an outlying uncertainty of the correlation parameters, which is not representative of the other~\nVelo regions, nevertheless they are shown in the final results.

\begin{table}[tp]
    \caption{
        \small 
        Systematic uncertainties on the~\becR and \becLambda~parameters. The~listed ranges correspond to the~lowest and highest values of the~given input determined across most of the~\nVelo bins in the~\pPb and \pbP samples~(see the~description in text for details). Negligible contributions are not listed. The~total uncertainty is a~quadratic sum of the~individual inputs.
        }
    \begin{center}
        \vspace*{-0.5cm}
        \begin{tabular}{lcccc}
    \hline
    & \multicolumn{2}{c}{\pPb dataset } & \multicolumn{2}{c}{\pbP dataset } \\
    Contribution & \sigSyst{\becR} [\myMath{\%}]  & \sigSyst{\becLambda} [\myMath{\%}] & \sigSyst{\becR} [\myMath{\%}] & \sigSyst{\becLambda} [\myMath{\%}] \\
    \hline
    Background scaling                      & \myMath{\myRangeInMath{ 4.5 }{ 9.0 }}        & \myMath{\myRangeInMath{ 3.5 }{ 11.0 }}    & \myMath{\myRangeInMath{ 4.5 }{ 6.5 }}      & \myMath{\myRangeInMath{ 3.0 }{ 9.5 }}     \\
    Background fit range                    & \myMath{\myRangeInMath{ 1.0 }{ 3.0 }}        & \myMath{\myRangeInMath{ 0.5 }{ 3.5 }}     & \myMath{\myRangeInMath{ 2.0 }{ 3.5 }}      & \myMath{\myRangeInMath{ 0.5 }{ 4.0 }}     \\
    Background fit -- fixed \becBkgMultZero & \myMath{\myRangeInMath{ 0.5 }{ 3.0 }}        & \myMath{\myRangeInMath{ 0.5 }{ 3.0 }}     & \myMath{< 0.5}                             & \myMath{< 0.5}                            \\
    Background fit -- resonances            & \myMath{\myRangeInMath{ 0.5 }{ 4.0 }}        & \myMath{\myRangeInMath{ 0.5 }{ 4.0 }}     & \myMath{\myRangeInMath{ 1.5 }{ 3.0 }}      & \myMath{\myRangeInMath{ 0.5 }{ 3.5 }}     \\
    \pid optimisation                       & \myMath{\myRangeInMath{ 0.5 }{ 1.5 }}        & \myMath{\myRangeInMath{ 0.5 }{ 5.0 }}     & \myMath{\myRangeInMath{ 0.5 }{ 10.5 }}     & \myMath{\myRangeInMath{ 0.5 }{ 8.5 }}     \\
    Fake tracks                             & \myMath{\myRangeInMath{ 0.5 }{ 5.5 }}        & \myMath{\myRangeInMath{ 1.0 }{ 8.0 }}     & \myMath{\myRangeInMath{ 0.5 }{ 4.5 }}      & \myMath{\myRangeInMath{ 0.5 }{ 8.0 }}     \\
    Requirement on \zPv                     & \myMath{\myRangeInMath{ 0.5 }{ 1.5 }}        & \myMath{\myRangeInMath{ 0.5 }{ 3.0 }}     & \myMath{\myRangeInMath{ 0.5 }{ 2.0 }}      & \myMath{\myRangeInMath{ 0.5 }{ 3.5 }}     \\
    Coulomb correction                      & \myMath{\myRangeInMath{ 0.5 }{ 1.5 }}        & \myMath{\myRangeInMath{ 1.0 }{ 2.5 }}     & \myMath{\myRangeInMath{ 0.5 }{ 2.0 }}      & \myMath{\myRangeInMath{ 0.5 }{ 3.0 }}     \\
    \like fit range (min)                   & \myMath{\myRangeInMath{ 1.5 }{ 5.0 }}        & \myMath{\myRangeInMath{ 1.0 }{ 8.5 }}     & \myMath{\myRangeInMath{ 0.5 }{ 3.5 }}      & \myMath{\myRangeInMath{ 0.5 }{ 5.5 }}     \\
    \like fit range (max)                   & \myMath{\myRangeInMath{ 0.5 }{ 1.0 }}        & \myMath{\myRangeInMath{ 0.5 }{ 2.0 }}     & \myMath{\myRangeInMath{ 0.5 }{ 2.0 }}      & \myMath{\myRangeInMath{ 0.5 }{ 3.0 }}     \\
    Reference sample                        & \myMath{\myRangeInMath{ 0.5 }{ 2.0 }}        & \myMath{\myRangeInMath{ 0.5 }{ 3.0 }}     & \myMath{\myRangeInMath{ 0.5 }{ 2.0 }}      & \myMath{\myRangeInMath{ 0.5 }{ 4.0 }}     \\
    \hline
    Total                                   & \myMath{\myRangeInMath{ 6.0 }{ 12.0 }}       & \myMath{\myRangeInMath{ 6.0 }{ 16.5 }}    & \myMath{\myRangeInMath{ 6.5 }{ 12.0 }}     & \myMath{\myRangeInMath{ 5.0 }{ 16.0 }}    \\
    \hline
\end{tabular}
    \end{center}
    \label{tab:pPb:systematics}
\end{table}

The leading source of systematic uncertainty is due to the parametrization of nonfemtoscopic background in the correlation function. It contains the effect related to the removal of the structures induced by two-body resonance decays from  the fits to the OS correlation function. The impact of the particular choice of those limits is investigated by repeating the analysis with the widths of the defined regions increased and decreased by 20\%. This value was already optimized in similar analyses performed by other experiments (\eg Ref.~\cite{Aaboud:2017}). Another effect in the determination of the cluster contribution is related to the choice of the range of the correlation function fits to the OS data, which is studied by varying its values within 10\%, leading to the similar range variation as for the SS fit. 

The impact of the $N_{0}$ value on the final correlation parameters is investigated by varying this value within $\sim$30\%, i.e. from 10 to 20. The chosen value represents a conservative approach, as the systematic uncertainty related to $N_{0}$ value is minor with respect to other sources. The scaling of the cluster contribution amplitude between the OS and SS pairs is found to be the dominant contribution to the systematic uncertainty, reaching up to 9\% (11\%) for the $R$ ($\lambda$) parameter. This was investigated by shifting the nominal parametrizations of the background scaling (see Eq.~\ref{eq:pPb:corrFunc:bkg:amplitude}) determined for the central results by $\pm$0.15 before using them in the final SS fits to investigate the influence of this procedure on the measured correlation parameters. Those values are chosen to comprise most of the individual results with the $z$ parameter left free in fits to the SS correlation function.

Systematic uncertainties related to the selection criteria involve the contribution related to the pion identification, which is based on the ProbNN(\pion) variable, by changing the requirement to increase the misidentified pions in the sample by $\sim$50\% with respect to the final selection. Another contribution is related to the misreconstructed tracks, which may degrade the purity of the selected pion sample and affect the final results. The misreconstructed tracks (mostly the clone ones) that could directly contribute to the SS pairs in the BEC-signal region are well controlled in the data (see Sec.~\ref{evtSel}), so no uncertainty is assigned to the clone tracks. A dedicated study is performed to evaluate the impact of additional fake tracks in the sample by modifying the selection requirement on the probability for a particle to be a fake track from $0.25$ to $0.50$ (which corresponds to the maximum value available in the dataset after the preselection). The fractions of fake tracks in the selected pion sample and of signal pairs containing a fake track (values determined using the simulation) for those two criteria are $\leq$1\% (see Table~\ref{tab:pPb:systematics:ghosts:fraction}).

\begin{table}[tp]
	\caption{
		\small 
    Fractions of fake tracks in the~selected pion sample and in the signal pairs~(with \becQ values restricted to the~\myMath{ \becQ < 1.0 \gev} region) containing a~fake track in the~\pPb and \pbP datasets. The~values are determined using the~simulation.
	}
	\begin{center}
    \vspace*{-0.5cm}
		\begin{tabular}{ccccc}
    \hline
    & \multicolumn{2}{c}{Single particle [\myMath{\%}]} & \multicolumn{2}{c}{Particle pair [\myMath{\%}]} \\
    Probability to be a fake track & \pPb & \pbP & \pPb & \pbP \\
    \hline
    \myMath{ < 0.25 } & \myMath{ 0.51 } & \myMath{ 0.43 } & \myMath{ 1.06 } & \myMath{ 0.81 } \\
    \myMath{ < 0.50 } & \myMath{ 0.57 } & \myMath{ 0.48 } & \myMath{ 1.19 } & \myMath{ 0.91 } \\
    \hline
\end{tabular}

	\end{center}
	\label{tab:pPb:systematics:ghosts:fraction}
\end{table}

The contributions of the fake tracks and the pion selection criteria optimization  to the systematic uncertainty are calculated as the absolute difference between the results obtained with the modified selection requirements and the central ones. In a limited number of bins, those inputs constitute the most important contributions to the total systematic uncertainty, together with the one related to the background scaling. 

Final-state Coulomb interactions for both the SS and OS pairs are taken into account in the fits to correlation functions.  A simple proportionality \mbox{\becREff = $\epsilon R$} is assumed~\cite{Aaboud:2017} and the $\epsilon$ values are varied between 0.5 and 2.0. This leads to the final values of \becREff corresponding to 0.5~fm and 8.0~fm, which are used to evaluate the systematic uncertainty related to the correction for Coulomb interactions.

The range and binning used in the fit to the correlation functions for SS pairs can affect the final results. The impact from the boundary in the low-$Q$ region is evaluated by altering it within 20\%, which corresponds to the values of 0.04 and 0.06~GeV. A similar procedure is implemented for the fit boundary at high-$Q$ values, where the modification at the level of 10\% is applied, leading to the upper fit range being limited to 1.8 and extended to 2.2~GeV. The smaller relative variation in the case of the upper fit range is motivated to stay within the range where the used parametrization describes correctly the effect related to the long-range correlations. Both contributions to the systematic uncertainty associated with the SS fit range are found to be relatively small. The systematic uncertainty related to the binning of the $Q$ variable in the correlation function is determined by doubling the bin width from the nominal  0.005~GeV to 0.010~GeV. The impact of this modification on the measured correlation parameters is negligible.

The construction of the reference sample is one of the basic aspects of the BEC analyses. The potential impact of the event-mixing implementation on the correlation parameters is assessed by varying the number of candidates available for the mixing, which is a parameter that can be tuned in the procedure. The nominal value of the number of candidates equal to 10 is changed to 50 and 100, and the analysis is repeated using the updated settings. The final contribution from the event-mixing to the systematic uncertainty is found to be small (see Table~\ref{tab:pPb:systematics}).

\section{Results}
\begin{figure}[tbp]
    \begin{center}
      \includegraphics[width=0.49\linewidth]{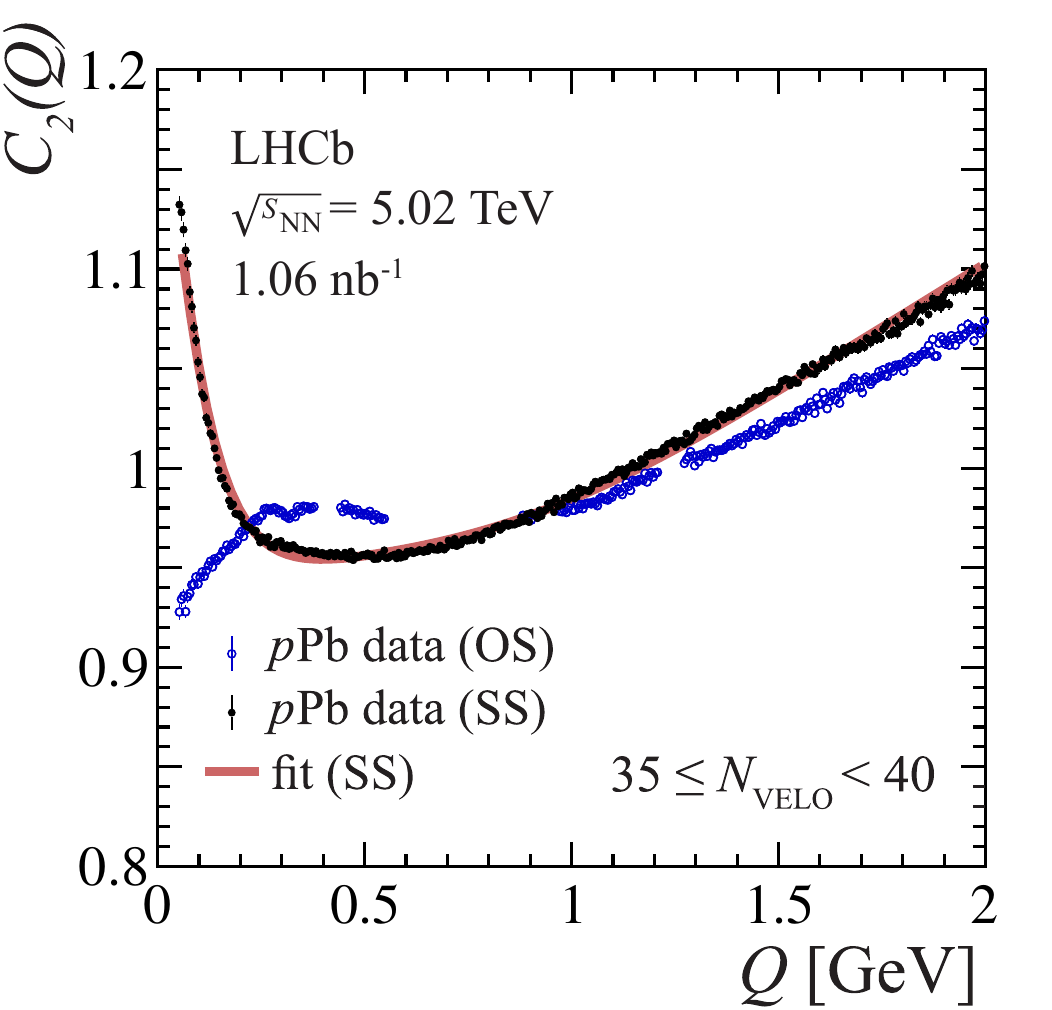}
      \includegraphics[width=0.49\linewidth]{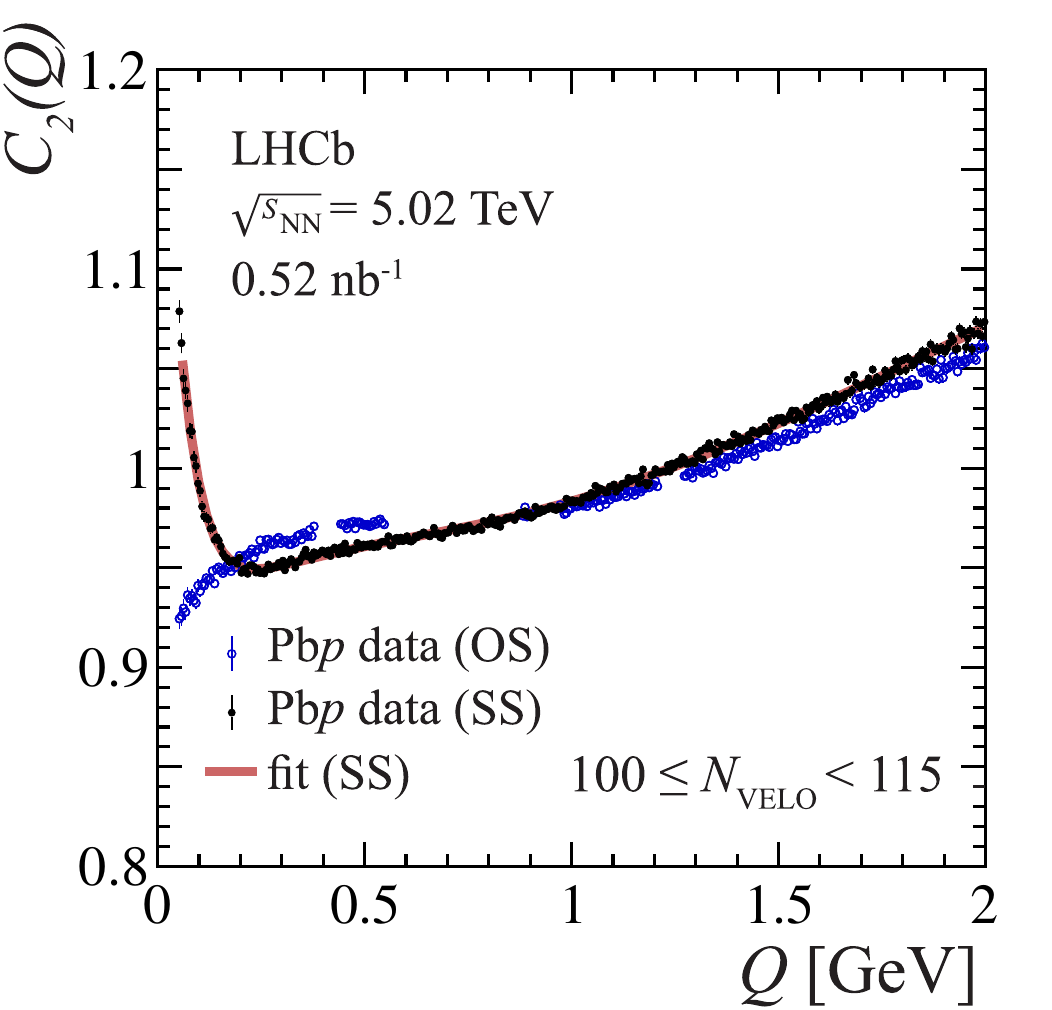}
      \vspace*{-0.5cm}
    \end{center}
    \caption{
      \small 
      Example of fits to the~\like \corrFuncFullPlural in (left)~a~\mbox{moderate-multiplicity} region~(\multTypical) of the~\pPb and (right)~\mbox{high-multiplicity} regime~(\multHigh) of the~\pbP dataset. The~black points correspond to the~\like \corrFuncFullPlural, while the~blue ones illustrate the~\unlike \corrFuncFullPlural which are used to estimate the~cluster contribution in the~given \nVelo bin. The~results of fits using Eq.~\ref{eq:pPb:corrFunc:like} are indicated by red solid lines. Only statistical uncertainties are shown.
    }	
    \label{fig:pPb:results:fits:like}
\end{figure}

\begin{figure}[tbp]
    \begin{center}
      \includegraphics[width=0.55\linewidth]{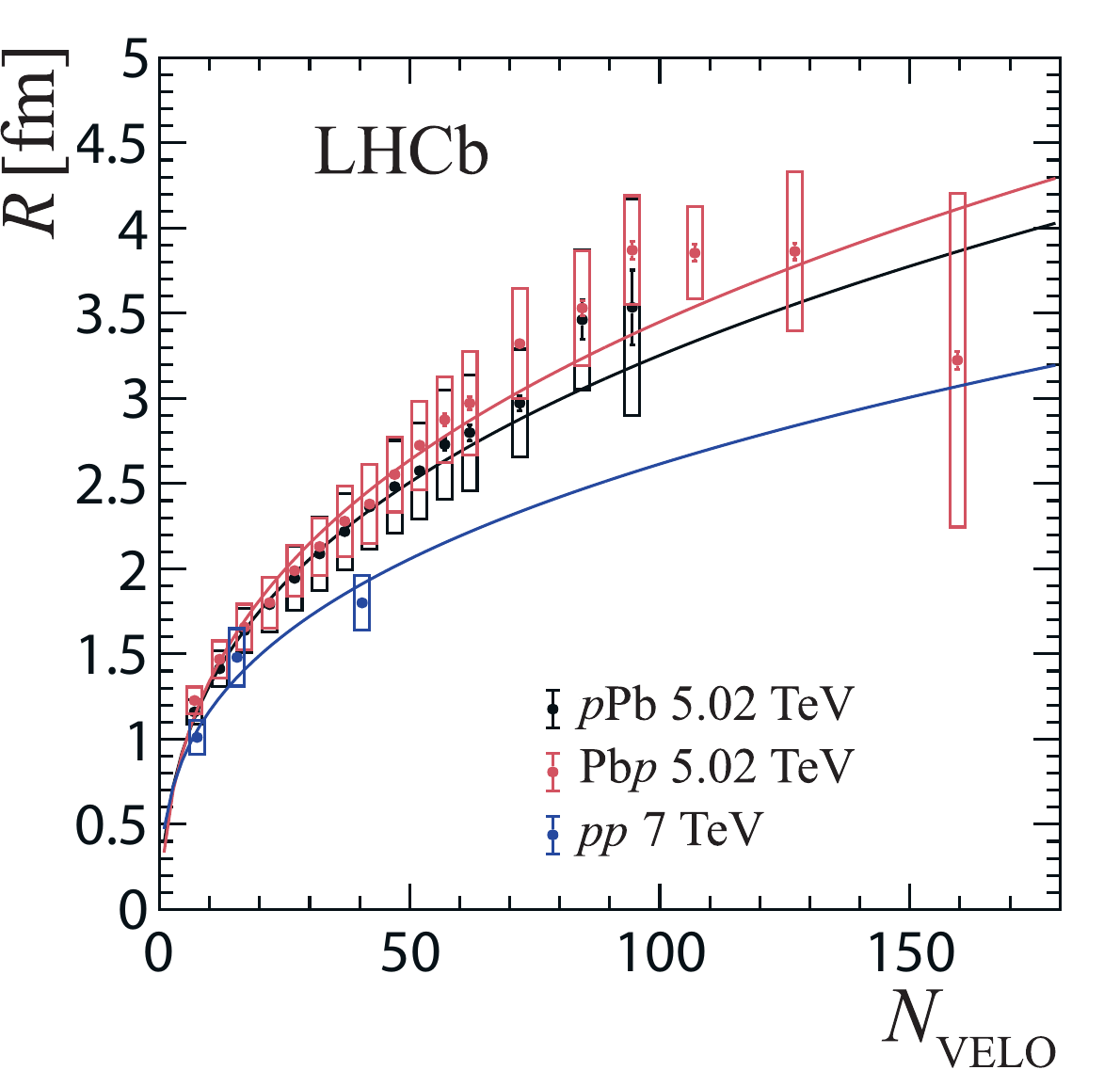}
      \vspace*{-0.5cm}
    \end{center}
    \caption{
      \small 
      Correlation radius as a~function of the~reconstructed \nChFull measured in the~\pp~\cite{LHCb-PAPER-2017-025}, \pPb and \pbP~collision systems in the~\lhcb experiment. Error bars indicate the~statistical uncertainties, while boxes illustrate the~systematic ones. Data~points are positioned at the~centres of the~multiplicity bins. Results of the~fits to the~observed radii scale linearly in the~cube root of the~reconstructed multiplicity (solid lines). 
    }	
    \label{fig:discussion:radius:systems}
\end{figure}

\begin{figure}[tbp]
    \begin{center}
      \includegraphics[width=0.55\linewidth]{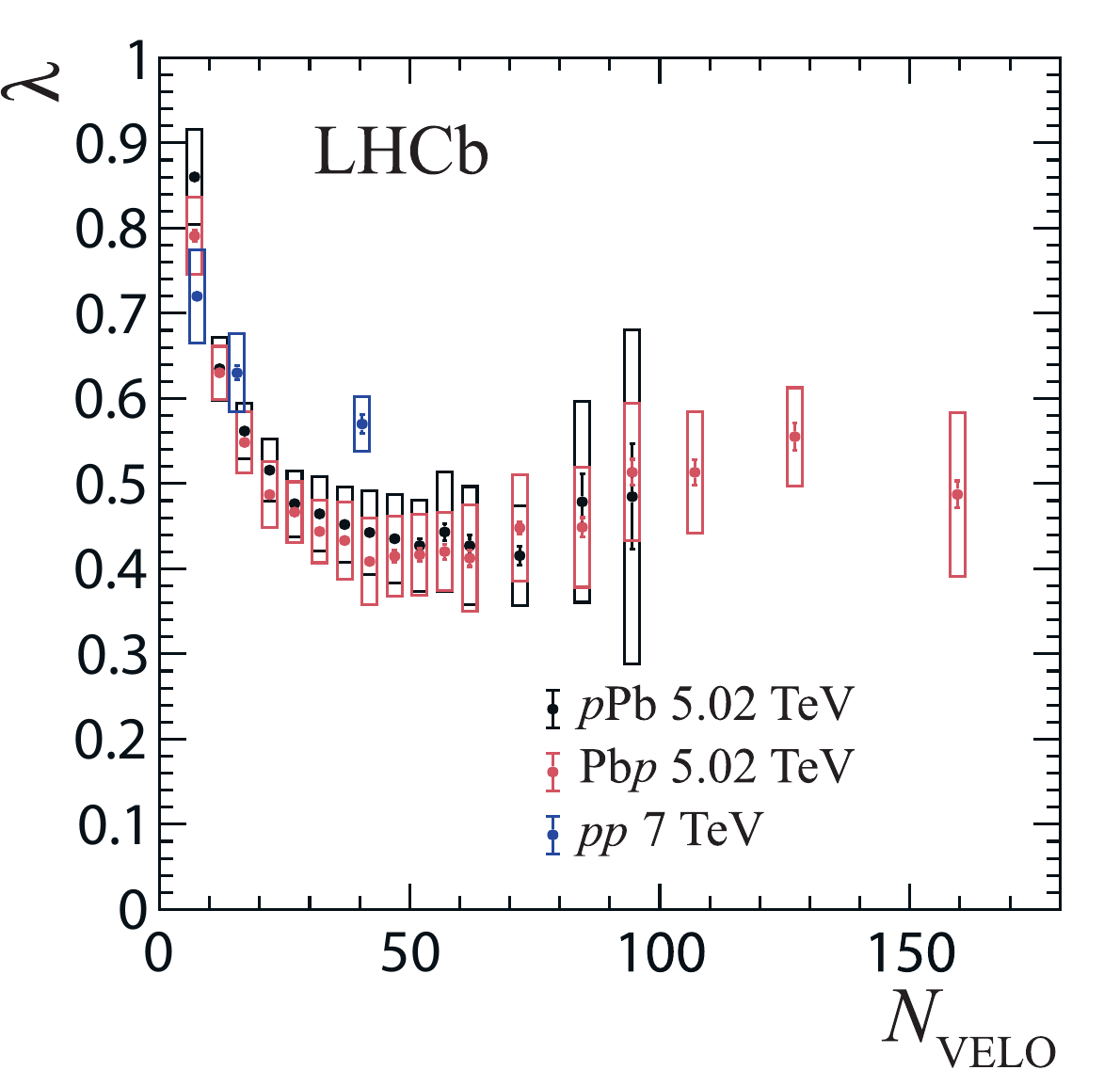}
      \vspace*{-0.5cm}
    \end{center}
    \caption{
      \small 
      Intercept parameter as a~function of the~reconstructed \nChFull measured in the~\pp~\cite{LHCb-PAPER-2017-025}, \pPb and \pbP~collision systems in the~\lhcb experiment. Error bars indicate the~statistical uncertainties, while boxes illustrate the~systematic ones. Data~points are positioned at the~centres of the~multiplicity bins.
    }	
    \label{fig:discussion:lambda:systems}
\end{figure}

The correlation parameters are determined by performing fits to the SS correlation functions in each individual $N_{\rm VELO}$ bin using Eq.~\ref{eq:pPb:corrFunc:like}. In this procedure, the parameters characterising the cluster contribution and the background scaling are fixed to the values measured in the previous steps of the analysis. The fits are performed in the full range of $Q$ variable (0.05-2.00~GeV) in the constructed correlation functions. Example results of the final fits to the SS correlation functions are presented in Fig.~\ref{fig:pPb:results:fits:like}. Correlation parameters determined from fits to the~\like \corrFuncFullPlural using Eq.~\ref{eq:pPb:corrFunc:like} in the~\nVelo bins for the~\pPb and \pbP datasets are presented in Table~\ref{tab:pPb:results:final}. The fit quality in BEC studies is not expected to be ideal due to various assumptions in the signal parametrization and the unknown theoretical parameterization of the nonfemtoscopic background effects. The results are complementary to the measurements performed at LHC energies in central rapidity regions~\cite{Adam:2015,Aaboud:2017,Sirunyan:2017}. The measured behaviour of the correlation parameters is compatible with observations from other experiments at LHC. In general, the correlation radius becomes larger with increasing event multiplicity, while the intercept parameter displays the opposite behaviour.  The determined $R$ ($\lambda$) parameters vary within 1-4~fm ($\sim$0.40-0.85) depending on the $N_{\rm VELO}$ interval. Correlation parameters determined in the BEC studies for the \pp~\cite{LHCb-PAPER-2017-025}, \pPb and \pbP collisions at LHCb are illustrated in Figs.~\ref{fig:discussion:radius:systems} and~\ref{fig:discussion:lambda:systems}. As it is observed in Fig.~\ref{fig:discussion:radius:systems} the measured correlation radii scale linearly with the cube root of the reconstructed charged-particle multiplicity. A simple fit illustrating this relationship is performed for different datasets (\pp, \pPb and \pbP). Only the statistical uncertainties of the measured $R$ values are taken into account in this fit. Similar scaling was also reported by other experiments at LHC for various collision systems~\cite{Adam:2015,Aaboud:2017,Sirunyan:2019}. It is a tendency compatible  with predictions of hydrodynamic models on the system evolution~\cite{Werner:2010,Bozek:2013,Shapoval:2013,Schenke:2014}. Although the results in both \pPb and \pbP samples agree well within the systematic uncertainties, it may be observed that the central $R$ values in the \pbP sample tend to be systematically higher than in the \pPb case, becoming more prominent with increasing multiplicity.

\newpage

\begin{table}[tp]
    \caption{
        \small 
        Correlation parameters determined from fits to the~\like \corrFuncFullPlural using Eq.~\ref{eq:pPb:corrFunc:like} in the~\nVelo bins for the~\pPb and~\pbP datasets. The first and second uncertainties are statistical and systematic, respectively.
        }
    \begin{center}
        \vspace*{-0.2cm}
        \footnotesize
        \begin{tabular}{ccccc}
    \hline
    & \multicolumn{2}{c}{\pPb dataset} & \multicolumn{2}{c}{\pbP dataset} \\
    \nVelo & \becR [\myMath{{\fm}}] & \becLambda & \becR [\myMath{{\fm}}] & \becLambda \\
    \hline
    \myMath{\myRangeInMath{5}{9}}    & \myMath{ 1.159 \pm 0.010 \pm 0.070 } & \myMath{ 0.860 \pm 0.006 \pm 0.056 } & \myMath{ 1.227 \pm 0.013 \pm 0.080 } & \myMath{ 0.791 \pm 0.007 \pm 0.045 } \\
    \myMath{\myRangeInMath{10}{14}}   & \myMath{ 1.413 \pm 0.010 \pm 0.105 } & \myMath{ 0.635 \pm 0.004 \pm 0.037 } & \myMath{ 1.469 \pm 0.013 \pm 0.108 } & \myMath{ 0.630 \pm 0.005 \pm 0.031 } \\
    \myMath{\myRangeInMath{15}{19}}   & \myMath{ 1.638 \pm 0.011 \pm 0.131 } & \myMath{ 0.562 \pm 0.004 \pm 0.033 } & \myMath{ 1.658 \pm 0.014 \pm 0.135 } & \myMath{ 0.548 \pm 0.005 \pm 0.036 } \\
    \myMath{\myRangeInMath{20}{24}}   & \myMath{ 1.790 \pm 0.011 \pm 0.161 } & \myMath{ 0.516 \pm 0.004 \pm 0.036 } & \myMath{ 1.801 \pm 0.015 \pm 0.148 } & \myMath{ 0.487 \pm 0.005 \pm 0.038 } \\
    \myMath{\myRangeInMath{25}{29}}   & \myMath{ 1.944 \pm 0.012 \pm 0.189 } & \myMath{ 0.476 \pm 0.004 \pm 0.039 } & \myMath{ 1.989 \pm 0.017 \pm 0.150 } & \myMath{ 0.467 \pm 0.005 \pm 0.036 } \\
    \myMath{\myRangeInMath{30}{34}}   & \myMath{ 2.088 \pm 0.014 \pm 0.214 } & \myMath{ 0.464 \pm 0.004 \pm 0.044 } & \myMath{ 2.130 \pm 0.019 \pm 0.169 } & \myMath{ 0.444 \pm 0.005 \pm 0.037 } \\
    \myMath{\myRangeInMath{35}{39}}   & \myMath{ 2.218 \pm 0.016 \pm 0.225 } & \myMath{ 0.452 \pm 0.005 \pm 0.044 } & \myMath{ 2.279 \pm 0.021 \pm 0.206 } & \myMath{ 0.433 \pm 0.006 \pm 0.045 } \\
    \myMath{\myRangeInMath{40}{44}}   & \myMath{ 2.364 \pm 0.019 \pm 0.250 } & \myMath{ 0.443 \pm 0.005 \pm 0.049 } & \myMath{ 2.380 \pm 0.024 \pm 0.233 } & \myMath{ 0.409 \pm 0.006 \pm 0.051 } \\
    \myMath{\myRangeInMath{45}{49}}   & \myMath{ 2.482 \pm 0.023 \pm 0.271 } & \myMath{ 0.435 \pm 0.006 \pm 0.052 } & \myMath{ 2.554 \pm 0.027 \pm 0.220 } & \myMath{ 0.415 \pm 0.007 \pm 0.047 } \\
    \myMath{\myRangeInMath{50}{54}}   & \myMath{ 2.575 \pm 0.028 \pm 0.281 } & \myMath{ 0.427 \pm 0.008 \pm 0.053 } & \myMath{ 2.725 \pm 0.031 \pm 0.259 } & \myMath{ 0.416 \pm 0.008 \pm 0.048 } \\
    \myMath{\myRangeInMath{55}{59}}   & \myMath{ 2.730 \pm 0.036 \pm 0.322 } & \myMath{ 0.443 \pm 0.010 \pm 0.070 } & \myMath{ 2.875 \pm 0.035 \pm 0.252 } & \myMath{ 0.420 \pm 0.009 \pm 0.046 } \\
    \myMath{\myRangeInMath{60}{64}}   & \myMath{ 2.799 \pm 0.046 \pm 0.341 } & \myMath{ 0.427 \pm 0.012 \pm 0.070 } & \myMath{ 2.972 \pm 0.040 \pm 0.306 } & \myMath{ 0.412 \pm 0.010 \pm 0.062 } \\
    \myMath{\myRangeInMath{65}{79}}   & \myMath{ 2.972 \pm 0.045 \pm 0.318 } & \myMath{ 0.415 \pm 0.011 \pm 0.059 } & \myMath{ 3.322 \pm 0.028 \pm 0.324 } & \myMath{ 0.448 \pm 0.007 \pm 0.062 } \\
    \myMath{\myRangeInMath{80}{89}}   & \myMath{ 3.462 \pm 0.115 \pm 0.410 } & \myMath{ 0.479 \pm 0.033 \pm 0.118 } & \myMath{ 3.531 \pm 0.043 \pm 0.337 } & \myMath{ 0.449 \pm 0.011 \pm 0.070 } \\
    \myMath{\myRangeInMath{90}{99}}  & \myMath{ 3.535 \pm 0.219 \pm 0.635 } & \myMath{ 0.485 \pm 0.062 \pm 0.196 } & \myMath{ 3.871 \pm 0.052 \pm 0.320 } & \myMath{ 0.513 \pm 0.015 \pm 0.081 } \\
    \myMath{\myRangeInMath{100}{114}} & --                                & --                                      & \myMath{ 3.854 \pm 0.049 \pm 0.270 } & \myMath{ 0.513 \pm 0.015 \pm 0.072 } \\
    \myMath{\myRangeInMath{115}{139}} & --                                & --                                      & \myMath{ 3.863 \pm 0.049 \pm 0.468 } & \myMath{ 0.555 \pm 0.016 \pm 0.057 } \\
    \myMath{\myRangeInMath{140}{179}} & --                                & --                                      & \myMath{ 3.225 \pm 0.053 \pm 0.979 } & \myMath{ 0.487 \pm 0.016 \pm 0.096 } \\
    \hline
\end{tabular} 
    \end{center}
    \label{tab:pPb:results:final}
\end{table}

\newpage

\newpage

\section{Conclusions}
Bose-Einstein correlations in pairs of same-sign charged pions in \pPb and \pbP collisions at LHCb are measured using a data-driven analysis method to account for effects related to the nonfemtoscopic background. The correlation parameters are determined in common intervals of VELO-track multiplicity. This measurement is the first of this type performed in the forward rapidity region at LHC energies. The correlation radius increases with the charged-particle multiplicity, while the intercept parameter tends to decrease in the region of lower charged-particle multiplicity. This trend is consistent with observations in the central rapidity region by other experiments at the LHC~\cite{Sirunyan:2017,Adam:2015a,Aaboud:2017}. The measured correlation radii scale linearly with the cube root of the charged-particle multiplicity, which is compatible with predictions based on hydrodynamic models~\cite{Werner:2010,Bozek:2013}. The proton-lead system is investigated both in the forward and backward directions due to asymmetric beams, and hints for a potential sensitivity of the correlation parameters to the rapidity are observed.

\newpage

\appendix
\section{Appendix. Distributions of the VELO-track multiplicity}
\label{app:mult}
\begin{figure}[bp]
	\begin{center}
	  \includegraphics[width=0.60\linewidth]{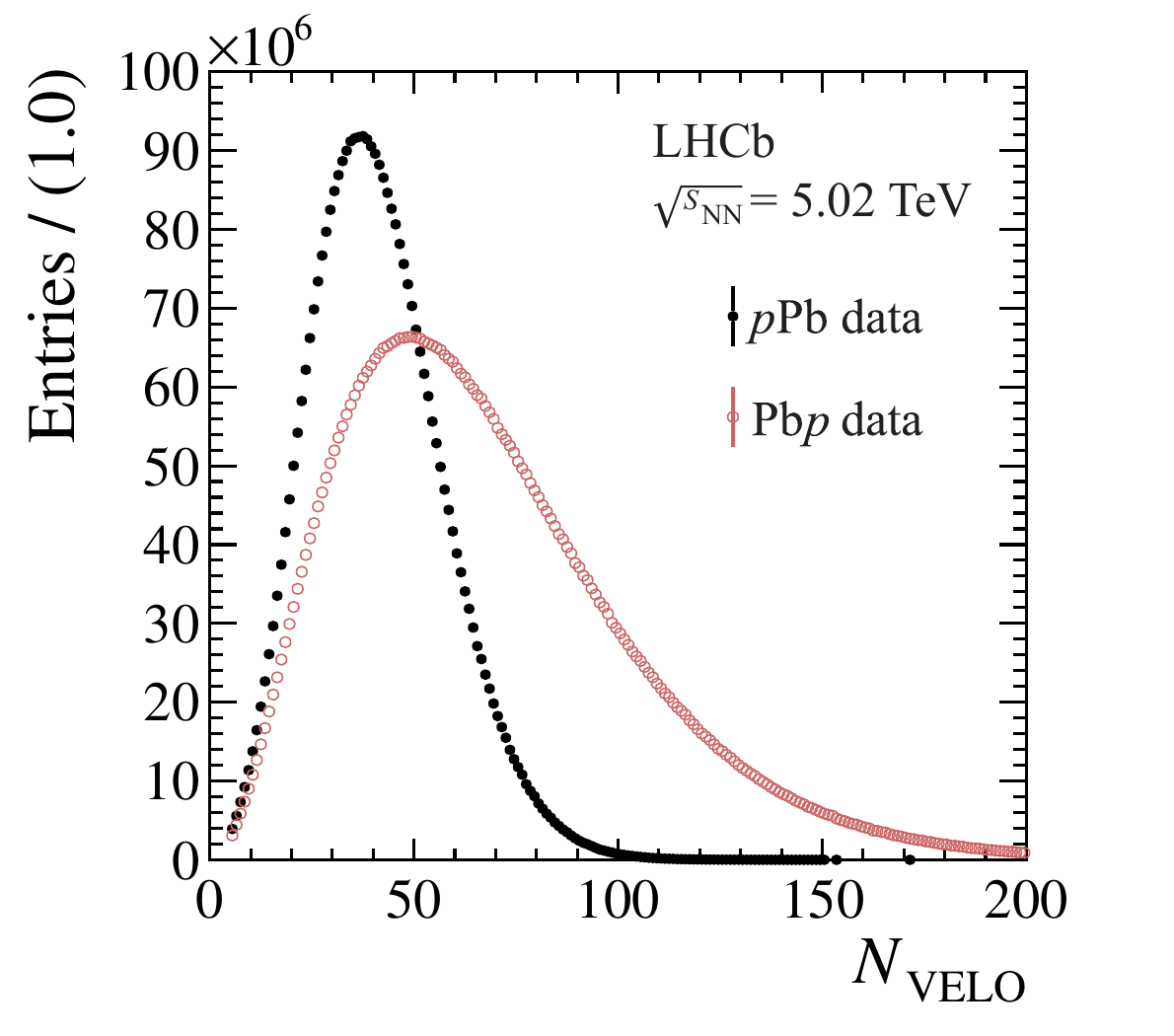}
	  \vspace*{-0.5cm}
	\end{center}
	\caption{
	  \small 
	  Distribution of the~selected signal pairs originating from primary vertices with the~given \nVeloFull in the~\pPb~(black dots) and \pbP~(red circles) data.
	}
	\label{fig:pPb:multiplicity}
\end{figure}

The~data samples are divided into bins of the~\nVeloFull, which is used as a~proxy for the~total \nChFull. The~division is optimized to obtain a~high number of bins with enough entries to perform the~measurement. This procedure is based on the~\nVelo distribution for the~signal pairs, which is shown in Fig.~\ref{fig:pPb:multiplicity} and allows selecting bins with similar signal yields for the~final analysis.

\newpage
\section*{Acknowledgements}
%
%
\noindent We express our gratitude to our colleagues in the CERN
accelerator departments for the excellent performance of the LHC. We
thank the technical and administrative staff at the LHCb
institutes.
We acknowledge support from CERN and from the national agencies:
CAPES, CNPq, FAPERJ and FINEP (Brazil); 
MOST and NSFC (China); 
CNRS/IN2P3 (France); 
BMBF, DFG and MPG (Germany); 
INFN (Italy); 
NWO (Netherlands); 
MNiSW and NCN (Poland); 
MEN/IFA (Romania); 
MICINN (Spain); 
SNSF and SER (Switzerland); 
NASU (Ukraine); 
STFC (United Kingdom); 
DOE NP and NSF (USA).
We acknowledge the computing resources that are provided by CERN, IN2P3
(France), KIT and DESY (Germany), INFN (Italy), SURF (Netherlands),
PIC (Spain), GridPP (United Kingdom), 
CSCS (Switzerland), IFIN-HH (Romania), CBPF (Brazil),
Polish WLCG  (Poland) and NERSC (USA).
We are indebted to the communities behind the multiple open-source
software packages on which we depend.
Individual groups or members have received support from
ARC and ARDC (Australia);
Minciencias (Colombia);
AvH Foundation (Germany);
EPLANET, Marie Sk\l{}odowska-Curie Actions, ERC and NextGenerationEU (European Union);
A*MIDEX, ANR, IPhU and Labex P2IO, and R\'{e}gion Auvergne-Rh\^{o}ne-Alpes (France);
Key Research Program of Frontier Sciences of CAS, CAS PIFI, CAS CCEPP, 
Fundamental Research Funds for the Central Universities, 
and Sci. \& Tech. Program of Guangzhou (China);
GVA, XuntaGal, GENCAT, Inditex, InTalent and Prog.~Atracci\'on Talento, CM (Spain);
SRC (Sweden);
the Leverhulme Trust, the Royal Society
 and UKRI (United Kingdom).




\newpage
\addcontentsline{toc}{section}{References}
\bibliographystyle{LHCb}
\bibliography{main,standard,LHCb-PAPER,LHCb-CONF,LHCb-DP,LHCb-TDR}

\newpage
\centerline
{\large\bf LHCb collaboration}
\begin
{flushleft}
\small
R.~Aaij$^{32}$\lhcborcid{0000-0003-0533-1952},
A.S.W.~Abdelmotteleb$^{51}$\lhcborcid{0000-0001-7905-0542},
C.~Abellan~Beteta$^{45}$,
F.~Abudin{\'e}n$^{51}$\lhcborcid{0000-0002-6737-3528},
T.~Ackernley$^{55}$\lhcborcid{0000-0002-5951-3498},
B.~Adeva$^{41}$\lhcborcid{0000-0001-9756-3712},
M.~Adinolfi$^{49}$\lhcborcid{0000-0002-1326-1264},
P.~Adlarson$^{77}$\lhcborcid{0000-0001-6280-3851},
H.~Afsharnia$^{9}$,
C.~Agapopoulou$^{43}$\lhcborcid{0000-0002-2368-0147},
C.A.~Aidala$^{78}$\lhcborcid{0000-0001-9540-4988}
% ...

{\footnotesize \it

% ...
$^{9}$Universit{\'e} Clermont Auvergne, CNRS/IN2P3, LPC, Clermont-Ferrand, France\\
% ...
$^{32}$Nikhef National Institute for Subatomic Physics, Amsterdam, Netherlands\\
% ...
$^{41}$Instituto Galego de F{\'\i}sica de Altas Enerx{\'\i}as (IGFAE), Universidade de Santiago de Compostela, Santiago de Compostela, Spain\\
% ...
$^{43}$European Organization for Nuclear Research (CERN), Geneva, Switzerland\\
% ...
$^{45}$Physik-Institut, Universit{\"a}t Z{\"u}rich, Z{\"u}rich, Switzerland\\
% ...
$^{49}$H.H. Wills Physics Laboratory, University of Bristol, Bristol, United Kingdom\\
% ...
$^{51}$Department of Physics, University of Warwick, Coventry, United Kingdom\\
% ...
$^{54}$School of Physics and Astronomy, University of Glasgow, Glasgow, United Kingdom\\
$^{55}$Oliver Lodge Laboratory, University of Liverpool, Liverpool, United Kingdom\\
% ...
$^{63}$Syracuse University, Syracuse, NY, United States\\
% ...
$^{77}$Department of Physics and Astronomy, Uppsala University, Uppsala, Sweden, associated to $^{54}$\\
$^{78}$University of Michigan, Ann Arbor, MI, United States, associated to $^{63}$\\
\bigskip
% ...
}
\end{flushleft}
%



\end{document}